\documentclass[aoas]{imsart}

\RequirePackage{amsthm,amsmath,amsfonts,amssymb}
\RequirePackage[authoryear]{natbib}
\RequirePackage[colorlinks,citecolor=blue,urlcolor=blue]{hyperref}%% uncomment this for coloring bibliography citations and linked URLs
\RequirePackage{graphicx}%% uncomment this for including figures
\usepackage{amsmath}
\usepackage{graphicx}
\usepackage{enumerate}
\usepackage{natbib}
\usepackage{url} % not crucial - just used below for the URL 
\usepackage{amssymb}
\usepackage{bbm}
\usepackage{xcolor}
\usepackage[colorlinks,citecolor=blue,urlcolor=blue]{hyperref}
\usepackage{multirow}
\usepackage{diagbox}
\usepackage{threeparttable}
\usepackage{caption}
\usepackage{subcaption}
\numberwithin{equation}{section}
\usepackage{comment}

\usepackage{algorithmicx}
\usepackage[ruled]{algorithm}
\usepackage{algpseudocode}
\newcommand{\f}{\frac}
\newcommand{\nn}{\nonumber}
\newcommand{\e}{\mathbbm{E}}
\renewcommand{\P}{\mathbb{P}}
\newcommand{\var}{\mbox{Var}}
\newcommand{\cov}{\mbox{Cov}}
 
\newtheorem{thm}{Theorem}[section]

\startlocaldefs
\endlocaldefs

\begin{document}

\begin{frontmatter}
%\title{\bf Local-area modeling of disease transmission using error-prone data with application to SARS-CoV-2}
\title{Early-Phase Local-Area Model for Pandemics Using Limited Data: A SARS-CoV-2 Application}
	%Robust modeling and inference of disease transmission using error-prone data with application to SARS-CoV-2}
%\title{A sample article title with some additional note\thanksref{T1}}
\runtitle{early-phase pandemic model}
%\thankstext{T1}{A sample of additional note to the title.}

\begin{aug}
%%%%%%%%%%%%%%%%%%%%%%%%%%%%%%%%%%%%%%%%%%%%%%%
%% Only one address is permitted per author. %%
%% Only division, organization and e-mail is %%
%% included in the address.                  %%
%% Additional information can be included in %%
\author[A]{\fnms{Jiasheng}~\snm{Shi}\ead[label=e1]{shijiasheng@cuhk.edu.cn}},
\author[C]{\fnms{Jeffrey}~\snm{Morris}\ead[label=e2]{Jeffrey.Morris@pennmedicine.upenn.edu}}
\author[B]{\fnms{David}~\snm{Rubin}\ead[label=e3]{Drubin@ucop.edu}}
\and
\author[C]{\fnms{Jing}~\snm{Huang}\ead[label=e4]{Jing14@Pennmedicine.upenn.edu}}
%  \author{
%      Jiasheng Shi \\
%Department of Biostatistics, Epidemiology and Informatics, \\University of Pennsylvania, Philadelphia, PA, USA\\ \\
%Jeffrey S. Morris \\
%Department of Biostatistics, Epidemiology and Informatics, \\University of Pennsylvania, Philadelphia, PA, USA\\ \\
%David M. Rubin \\
%PolicyLab, Children's Hospital of Philadelphia, Philadelphia, PA, USA\\ \\
%    and \\ \\
%  Jing Huang \\
%Department of Biostatistics, Epidemiology and Informatics, \\University of Pennsylvania, Philadelphia, PA, USA\\
%}

%%%%%%%%%%%%%%%%%%%%%%%%%%%%%%%%%%%%%%%%%%%%%%
%% Addresses                                %%
%%%%%%%%%%%%%%%%%%%%%%%%%%%%%%%%%%%%%%%%%%%%%%
\address[A]{School of Data Science, The Chinese University of Hong Kong, Shenzhen,
	China.\printead[presep={,\ }]{e1}}

\address[C]{Department of Biostatistics, Epidemiology and Informatics, University of Pennsylvania\printead[presep={,\ }]{e2,e4}}

\address[B]{University of California \printead[presep={,\ }]{e3}}

\end{aug}

\begin{abstract}
The emergence of novel infectious agents presents challenges to statistical models of disease transmission. These challenges arise from limited, poor-quality data and an incomplete understanding of the agent. Moreover, outbreaks manifest differently across regions due to various factors, making it imperative for models to factor in regional specifics. In this work, we offer a model that effectively utilizes constrained data resources to estimate disease transmission rates at the local level, especially during the early outbreak phase when primarily  infection counts and aggregated local characteristics are accessible. This model merges a pathogen transmission methodology based on daily infection numbers with regression techniques, drawing correlations between disease transmission and local-area factors, such as demographics, health policies, behavior, and even climate, to estimate and forecast daily infections. We incorporate the quasi-score method and an error term to navigate potential data concerns and mistaken assumptions. Additionally, we introduce an online estimator that facilitates real-time data updates, complemented by an iterative algorithm for parameter estimation. This approach facilitates real-time analysis of disease transmission when data quality is suboptimal and knowledge of the infectious pathogen is limited. It is particularly useful in the early stages of outbreaks, providing support for local decision-making.
\end{abstract}


\end{frontmatter}

\section{Introduction}
\label{introduction}

When a novel infectious agent emerges in a population lacking prior immunity, it can spread quickly, posing a threat of national or global outbreaks. An immediate response is crucial, but data infrastructures often fail to capture all the necessary data during the initial phase of an outbreak. For example, during the COVID-19 pandemic, although the CDC began collaborating with state, tribal, local, and territorial health departments in January 2020 to collect and validate COVID-19 case and death data, an automated data transfer system using an application programming interface was not fully established until July to improve data collection and quality at the jurisdiction level. It was only in December 2020 that the CDC set up business rules and curated official online data sources for most US counties (\citealp{khan2022improving,khan2023tracking}). Furthermore, the first data collection on SARS-CoV-2 seroprevalence in the US did not begin until July 2020, and the estimates were not available until September 2021, with estimates only at the national level (\citealp{jones2021estimated, jones2023estimates}). Consequently, policymakers faced challenges in making timely, informed decisions with limited and potentially unreliable data in the pandemic's early stages. 

On the other hand, it's important to note that while only a limited spectrum and quality of disease data are available, local factors may be harnessed to model an outbreak's emergence and predict its trajectory (\citealp{pica2012environmental, stewart2014spatiotemporal, rubin2020association, alam2021influences}). Moreover, outbreaks are often driven by local transmission events, which can vary significantly from region to region. Local-area disease modeling is likely to be more useful than national modeling for policymakers to design mitigation strategies that consider regional-specific characteristics and needs.

A specific example is the policy questions posed by local governors during the initial stages of the COVID-19 pandemic in March 2020. At that time, governments deliberated over whether to enforce a lockdown and, if so, for how long. Some proposed a stringent short-term strategy, advocating for a strict lockdown until summer to entirely contain the virus, assuming that the virus would not survive in high temperatures. In contrast, others recommended a milder but more extended approach to social distancing, suggesting that temperature might not inhibit the virus's spread, warranting a longer-term strategy \citep{john_2020,allison_2020,james_2020}. To make an informed decision, it was essential to understand the impact of temperature on the virus's transmission rate and to compare the effects of temperature and social distancing on the spread of this new pathogen. At that time, the pandemic had only begun a few weeks prior, so there wasn't sufficient data within a single location to estimate the effect of temperature. A national model would also not be able to answer such a local question. We proposed utilizing data from multiple counties, spanning different latitudes with varying temperatures experienced during those initial weeks, to help gauge the impact of temperature, while adjusting for county-level covariates.

Regression models are often well-suited statistical tools for such an analysis. The challenge lies in creating or adapting an epidemiological model that captures the dynamics of disease transmission using limited and potentially unreliable data. This model must also be harmoniously integrated with a regression model to gauge the impact of local factors on the transmission process. Not all models are suitable for this task. For example, classical compartment models, which segregate individuals within a population into distinct compartments over varying time intervals, present challenges when data for all compartments are not available. The more compartments assumed, the more data variables are needed to fit the model. Additionally, they also face challenges when attempting to integrate with regression models. These models often treat the disease transmission parameter as a piecewise constant function over time, and they rely on solving a complex set of differential equations, either deterministic or stochastic, to determine disease transmission rates. In contrast, the time-since-infection (TSI) models, rooted in assumptions about the renewal process of case incidence numbers (which are treated as random variables and might be the sole disease data variables required for model fitting), can be tailored to regression models by applying specific distributional assumptions to the random variables representing case incidence numbers (\citealp{quick2021regression}).

The TSI models are designed around the premise that the number of new infections at a particular moment, such as day $t$, is influenced by three primary factors: the count of recent infections, the reproduction number at that moment, and an infectiousness function that measures the contagiousness of an infected person at each day since infection. For illustration, let 
$I_{it}$ represent the new infections on day $t$ in location $i$. If we know the infections before day $t$, denoted by $I_{i0}, ..., I_{i,t-1}$, the anticipated new infections on day $t$ in location $i$ can be deduced as the multiplication of the instantaneous reproduction number and the infection potential on that particular day at the given location. This relationship can be expressed as:
\begin{align}\label{eqn:it}
E(I_{it}| I_{i0},\cdots,I_{i,t-1}) = R_{it} \Lambda_{it},
\end{align}
where $R_{it}$ represents the time-varying reproduction number and $\Lambda_{it} \triangleq \sum_{s=1}^t I_{i,t-s}\omega_s$ stands for the infection potential on day $t$ in location $i$. The infection potential is shaped by the current number of infectious individuals and the infectiousness function $\omega_s$, which quantifies the infectiousness of the existing infectious individuals on the $s$-th day after infection, with $\sum_{s=0}^{\infty} \omega_s =1$. This infectiousness function can be approximated using the probability distribution of the serial interval or generation time (\citealp{svensson2007note}). To estimate $R_{it}$, one may assume distribution assumptions for $I_{it}$ based on equation (\ref{eqn:it}) and use methods like maximum likelihood or Bayesian approaches (\citealp{cori2013new,who2014ebola,quick2021regression}) with a predetermined $\omega_s$. Due to the empirical nature of this model, it can be readily adapted to regression models, broadening its applicability (\citealp{quick2021regression}). 

TSI models were extensively used to analyze the COVID-19 pandemic  \citep{pan2020association,nouvellet2021reduction,amman2022viral,nash2022real,ge2023effects}, particularly in the early stages, likely because data on reported daily cases were mainly available during that time period. The models demonstrated superior accuracy in estimating disease transmission compared to many other methods \citep{gostic2020practical,nash2022real}. However, estimates of $R_{it}$ can be biased due to data errors or incompleteness in $I_{it}$, which can stem from batch reporting, delayed reporting, under-ascertainment, manual errors, and other factors.. There are different ways to address these data challenges, one of which is to leverage additional data sources. For example, recent work have used serological studies and symptom surveys to address the under-ascertainment and delayed reporting of case incidence  (\citealp{lison2023generative,quick2021regression,noh2021estimation,dempsey2020addressing}). Relevant to our proposed work, the model introduced by \citet{quick2021regression} integrates testing data and population-based serological surveys to account for under-ascertainment and delayed reporting. However, these data are often unavailable at the local-area level and in the early stages of outbreaks.

In our study, we address these challenges in a distinct context, focusing on the early phase of an outbreak when additional data resources might be unavailable at the local-level. Considering the scarce of disease data and limited confidence about the pandemic models, our model introduces a general measurement error term, instead of relying on additional data and parametric assumptions, to model the mechanism of errors. This term accounts for deviations in $R_{it}$ that may arise from data inaccuracies or inaccurate assumptions about the model and cannot be explained by observed local-area factors. In essence, the proposed model is a local-area disease transmission model robust to data errors, merging a TSI model of transmission dynamics with regression models. This framework identifies key local-area factors influencing disease transmission variability across regions. We utilize the quasi-score method to ease distributional assumptions about the data and make minimal parametric assumptions on the measurement error component, bolstering model robustness. Our model offers flexibility in estimating the $R_{it}$ values even when relying on imperfect data or potentially mis-specified model assumptions. This is especially valuable during an outbreak's early stages when additional data resources like testing data and serological surveys aren't available and knowledge about the infectious agent is sparse. Furthermore, we introduce an online estimator coupled with an efficient iterative algorithm to estimate model parameters. This methodology supports continuous monitoring and dynamic forecasting of disease transmission under various scenarios, offering critical insights for local decision-making.

%The rest of the paper is organized as follows. We introduce the proposed model in Section \ref{main model}, followed by the asymptotic results and property of estimating procedure in Section \ref{properties}. We show a robust performance of our method under various simulation scenarios in Section \ref{simulation}, an application of our model to the early stage COVID-19 data in Section \ref{real data}, and a miscellany of discussions in Section \ref{discussion}. Proofs for theoretical properties are shown in the Supplementary Materials.

\section{Method} \label{main model} In this section, we introduce the proposed model and estimation procedure, as well as the approach to construct confidence intervals for the estimates.

\subsection{An early phase local-area model for modeling disease transmission}
\label{section without measurement error}
Building on Equation (\ref{eqn:it}), for counties $i=1,\cdots,n$, and time points $t = 1,\cdots, N$, we posit moment constraints on the number of new infections on day $t$ in location $i$ given previously reported existing infections prior to day $t$, as:
\begin{align} \label{general GLM setting}
\mu_{it} &\triangleq \e \big( I_{it} | \mathcal{F}_{i,t-1} \big) = \e \big( R_{it} | \mathcal{F}_{i,t-1} \big) \Lambda_{it},\qquad
\nu_{it} \triangleq \var \big( I_{it} | \mathcal{F}_{i,t-1} \big)=g(\mu_{it})\cdot \phi_i,
\end{align}
where $\mathcal{F}_{i,t}$ is the filtration of past incident cases information, $\phi_i$ is a dispersion parameter, $g(\cdot)$ is a known variance function, and $\mu_{it}=R_{it} \Lambda_{it}$ when $R_{it} \in \mathcal{F}_{i,t-1}$. When $g(\cdot)$ is unknown, smoothing techniques, e.g. local polynomial fitting, can be used to estimate the variance function \citep{chiou1998quasi}. More discussions on unknown variance functions can be found in Section A of the Supplementary Materials.

As previously noted, $\Lambda_{it} \triangleq \sum_{s=1}^t I_{i,t-s}\omega_s$ represents the infection potential, determined by the current number of infectious individuals and the infectiousness function $\omega_s$. We assume that the $\omega_s$'s represent the probability distribution of the serial interval or generation time, and set $\omega_s=0$ when $s =0$ or $s>\eta$, with $\eta$ denoting the duration from infection to either recovery or mandatory quarantine. We use $\omega=\{ \omega_s, 1\leq s\leq \eta \}$ to denote the vector of $\omega_s$. In practice, $\omega$ is a vector of positive values, often obtained from epidemiological studies of infectious pathogens \citep{he2020temporal}. In cases where multiple studies on serial intervals exist, a meta-analysis can be conducted to obtain a pooled estimate.

To estimate $R_{it}$ and investigate the impact of local area factors on disease transmission, we further assume a time series model as follows:
\begin{align}
h(R_{it}) &=  W^T_{it} \check{\beta}_i +X_{it}^T \check{\beta}_0 +  \sum_{m=1}^{q} \theta_m f_m (D_{1,it},D_{2,it},D_{3,it}) +\epsilon_{it},   \label{model_assumption measurement error} 
\end{align}
where $D_{1,it} =\big\{ X_{ij}    \big\}_{0\leq j\leq t}$,  $ D_{2,it}=\big\{ I_{ij}  \big\}_{0\leq j\leq t-1}$, $D_{3,it}=\big\{ R_{ij}  \big\}_{0\leq j\leq t-1}$, and $\mathcal{F}_{i,t-1} =\sigma(D_{1,it}\cup D_{2,it})$. The term $\sum_{m=1}^{q} \theta_m f_m (D_{1,it},D_{2,it},D_{3,it})$ describes the time-series dependency of the disease transmission rate. Specifically, $f_m(\cdot)$ is a function of past data and $\theta_m \geq 0$ for $1\leq m < q$. $f_{q} \equiv 1$ is set as the constant function and $\theta_{q}$ is the intercept term. The $h(\cdot)$ is a known link function. Moreover, $W_{it}\in \mathbb{R}^{p_1}$ and $X_{it} \in \mathbb{R}^{p-p_1}$ are vectors of local-area factors for location $i$ on day $t$ associated with disease transmission. The factors in $W_{it}^T$ have fixed location-specific effects, while the factors in $X_{it}^T$ have common effects across locations. To simplify notion, we consolidate the presentation of all local-area factors as $Z_{it} = (X_{it}^T, \overbrace{0, \cdots,0}^{p(i-1)} , W_{it}^T,  \overbrace{0, \cdots,  0}^{p(n-i)} )^T$, present the vector of all regression coefficients of these factor as $\beta = (\check{\beta}_0^T,\check{\beta}_1^T,\cdots, \check{\beta}_n^T)^T\in \mathbb{R}^p$, and denote the vector of all other regression coefficients as $\theta = (\theta_1,\cdots,\theta_{q})^T\in \mathbb{R}^{q}$.

In practice, a choice for Equation (\ref{model_assumption measurement error}) can be a non-stationary log-transmission model with an autoregressive process as follows:
\begin{equation}
\log(R_{it})= Z_{it}^T \beta +  \sum_{m=1}^{q-1} \theta_m   \log (R_{i,t-m}) + \theta_{q}+\epsilon_{it}, \;  \Vert \theta \Vert_1 <1 \;\text{and}\; \mu_{it} =\nu_{it}, \;{\rm for}\; t\geq q, \;  \theta_m\geq 0. \label{non-stationary special case}
\end{equation}

In the above models, to counter potential biases arising from data errors and incorrectly specified model assumptions, we incorporate $\epsilon_{it}$ as a measurement error term for $R_{it}$. This term can also be viewed as a random effect term of $R_{it}$, capturing unexplained deviations in $R_{it}$ that could result from inaccuracies in reported infection data or model mis-specifications. At this point, our proposed model breaks down the serial dependency of incidence cases into two components: Equation (\ref{general GLM setting}) encapsulates the serial dependency inherent in the generative nature of disease transmission given $R_{it}$, and Equation (\ref{model_assumption measurement error}) portrays the serial dependency arising from the $R_{it}$ interdependence between adjacent time points, elucidated by covariates and time series structures. %Conceptually, our proposed model echoes the characteristics of a recurrent neural network, as depicted in Figure \ref{workflow}.

%\begin{figure}[htbp!]
%	%	\begin{minipage}{0.45\linewidth}
%	\centering	
%	\includegraphics[width=0.93\linewidth]{pics//QSOEID_JASA_version1.jpg}
%	\caption{The proposed method, described in Equations (\ref{general GLM setting}) and (\ref{model_assumption measurement error}) for a specific location, is illustrated as a recurrent neural network's computational graph. We remove the location subscript $i$ in the figure to simplify notation. By combining the data of incident cases ($I_t$) and local-area factors ($X_t$), the proposed model constructs and solves an estimating function (the loss function, $L_t$) to estimate disease transmission ($R_t$) and the association between local-area factors and disease transmission (the model output $O_t$).  }
%	\vskip -2em
%	\label{workflow}
%\end{figure}
%\vspace{5mm}

%\noindent where $\mu_{it}=\nu_{it}$. In this model, an autoregressive $AR(q)$ model is used to model the time series dependency and the non-stationary of $R_{it}$ is due to the variation of exogenous covariates. We impose the regularity condition $\Vert \theta \Vert_1 <1$ to ensure the causality of the time series when exogenous terms are bounded and known. The parameter $\beta_i$ quantifies the association between local-area factors and disease transmission. %One could also allow $\beta$ to be location-specific, using random effects models to allow the associations to be heterogenous across locations.

\subsection{Estimation}
With the measurement error term in Equation (\ref{model_assumption measurement error}), the estimation of $R_{it}$ and parameters in the proposed model becomes difficult. In a simple scenario when $\epsilon_{it}$ is assumed to be 0, Equation (\ref{model_assumption measurement error}) reduces to 
\begin{align}
h(R_{it}) &=  Z_{it}^T {\beta}+ \sum_{m=1}^{q} \theta_m f_m (D_{1,{it}},D_{2,it},D_{3,it}).  \label{model_assumption} 
% \log(R_{it})-X_t^T \beta &=   \sum_{i=1}^q \theta_i  \left[  \log (R_{t-j}) -x_{t-j}^T \beta   \right] +\epsilon_t = \log (\tilde{\epsilon}_t )
\end{align}%which is similar to the model proposed in \cite{zeger1988markov}. 
In this situation, an estimator of the parameters $\gamma=(\beta^T,\theta^T)^T$, denoted as $\hat{\gamma}$, can be obtained by solving $U_N({\gamma})=0$, where $U_N(\gamma)$ is the quasi-score function 
\begin{equation}
U_N(\gamma)= \sum_{i=1}^{n} \sum_{t=q}^{N} \big(  \frac{ \partial \mu_{it}}{\partial \gamma} \big)^T \nu_{it}^{-1} (I_{it}- \mu_{it})\triangleq   \sum_{i=1}^{n} \sum_{t=q}^N \xi_{it} (\gamma).   \label{quasi-score equation}
\end{equation}

However, when $\epsilon_{it}$ is a non-degenerate random variable, calculating the expectation in $\mu_{it}=\e [R_{it} |\mathcal{F}_{i,t-1}]\Lambda_{it}$ and solving $U_N(\gamma)=0$ are generally difficult. Similar models were studied by \cite{davis2003observation} to model count data using the log link function, where the measurement error was assumed to be a stationary process and calculated using an autoregressive moving average recursion with a distributed lag structure. Here, given that $\{R_{it}\}_{1\leq t\leq N}^{1\leq i\leq n}$ are not directly observed, we describe an iterative algorithm for estimating parameters in (\ref{model_assumption measurement error}) without requiring any stationary structure and distribution assumption on the measurement error term. Briefly, the proposed iterative algorithm consists of two steps. The first step uses the second layer of the model, the time series model, to construct a semi-parametric, locally efficient estimator of $\beta$ based on a location-shift regression model. In this step, $\beta$ and $R_{it}$ can be written as functions of $\theta$ using an initial estimate of $\theta$ or an estimate from the last iteration. In the second step, the parameters were updated using the first layer of the model, the quasi-score function. Specifically, for $t=q,\cdots,N$, $i=1,\cdots,n$, we first rearrange Equation (\ref{model_assumption measurement error}) to
\begin{equation*}
h(R_{it})-\sum_{m=1}^{q} \theta_m f_m (D_{1,it},D_{2,it},D_{3,it})= Z_{it}^T \beta +\epsilon_{it}.
\end{equation*}

\noindent Assuming the Gauss-Markov assumption \citep{wooldridge2016introductory}, i.e., the conditional independence of $\{ \epsilon_{it}|Z_{it}, t\geq 1 \}$, $\e [\epsilon_{it}|Z_{it} ]=0$ and homoscedasticity, the above equation forms a location-shift regression model. Thus, assuming $\theta$ is known, a semi-parametric locally efficient estimator \citep{tsiatis2007semiparametric} for $\beta$ is obtained by
\begin{align}
\hat{\beta}&= \left\{  \sum_{i=1}^n \sum_{t=1}^N (Z_{it}-\bar{Z})(Z_{it} -\bar{Z})^T        \right\}^{-1} \label{sudoestimator} \\
& \qquad \qquad  \sum_{i=1}^n \sum_{t=1}^N (Z_{it} -\bar{Z}) \left(   h(R_{it})- \sum_{m=1}^q \theta_m f_m (D_{1,it},D_{2,it},D_{3,it})   \right), \nn
\end{align}
where $\bar{Z}=\big(   \sum_{i=1}^n \sum_{t=1}^T Z_{it}  \big)/Nn$, $N$ is the total number of observed time points  and $n$ is the number of counties. Based on this formulation, we develop an iterative estimation method, namely the \underline{local}-area disease transmission model using the \underline{qu}asi-score \underline{est}imation (LOCAL-QUEST), to estimate $\gamma$ and $R_{it}$ by iteratively updating the locally-efficient semi-parametric estimator of $\beta$ and the quasi-score estimator of $\theta$ as shown in {\it Algorithm 1}. Intuitively, with an initial estimate of $\theta$ and $\beta$, intermediate semi-parametric locally efficient estimator of $\beta$ can be written as a function of $\theta$ and these initial estimates using equations ({\ref{sudoestimator}}) and (\ref{model_assumption}). Then $\hat\theta$ can be updated by solving equation (\ref{quasi-score equation}) and $\hat\beta$ is updated by the updated $\hat\theta$ and equation ({\ref{sudoestimator}}). The algorithm is then proceeded iteratively until it converges. Due to its iterative nature, this procedure provides online estimators \citep{bottou1998online,farhang2013adaptive} for $\{ R_{it} \}_{1\leq t\leq N}^{1\leq i\leq n}$ and $\gamma$, allowing daily updates or update when new data become available. This can be particularly useful for real-time monitoring of outbreaks, as disease incidence and local-area factors are changing over time, and the data stream comes in continuously. The parameters are updated incrementally as new data arrives, which also reduces the requirement of computational memory. 

\alglanguage{pseudocode}
\begin{algorithm}[htbp!] \label{algorithm}
	\small
	\caption{: The LOCAL-QUEST algorithm.\\Estimation of the instantaneous reproduction number and local-area effects}
	\label{Algorithm:Estimation of instantaneous reproduction number and infectiousness profile}
	\begin{algorithmic}[1]
		\Procedure{}{}
		Initialize the parameters using data from a small initial period of the pandemic $\tau_0$ and the model (\ref{model_assumption}) that ignores measurement error. Denote the initial values as $\hat\theta^{(\tau_0)}$, $\hat\beta^{(\tau_0)}$ and $ \{\hat{R}_{it}^{(\tau_0)}\}_{0 < t\leq \tau_0}^{1\leq i\leq n}$. 
		\Loop
		\For {$k > \tau_0$, \textbf{and} $k\leq N$, \textbf{with estimates} $\hat\theta^{(k-1)}, \hat\beta^{(k-1)}$, \textbf{and} $\{\hat{R}_{it}^{(k-1)}\}_{0 < t\leq k-1}^{1\leq i\leq n}$ \textbf{obtained in the last iteration,}}
		\State \textbf{write the approximations of $\hat\beta$ and $\{\hat R_{it}\}_{\tau_0+1\leq t\leq k}^{1\leq i\leq n}$ as a function of $\theta$ using equations $(\ref{sudoestimator})$ and $(\ref{model_assumption})$, that is}
		\begin{align}
		\tilde{\beta}^{(k)}&\triangleq \left\{ \sum_{i=1}^{n} \sum_{t=1}^{k-1} (Z_{it}-\bar{Z}_{k-1})(Z_{it} -\bar{Z}_{k-1})^T        \right\}^{-1}  \label{betatilde} \\
		&\qquad \times \sum_{i=1}^{n} \sum_{t=1}^{(k-1)} (Z_{it} -\bar{Z}_{k-1}) \Big[ h(\hat{R}_{it}^{(k-1)}) - \sum_{m=1}^q \theta_m f_m \big(D_{1,it},D_{2,it},\hat{D}_{3,it}^{(k-1)} \big) \Big] \nn \\
		{\text {and }}	&\tilde{R}_{it}^{(k)} \triangleq h^{-1} \left( \sum_{m=1}^q \theta_m f_m (D_{1,it},D_{2,it},\hat{D}_{3,it}^{(k-1)}) + Z_{it}^T \tilde{\beta}^{(k)} \right),\label{tildeY} 
		\end{align}
		\State \textbf{Obtain $\hat\theta^{(k)}$ by solving the quasi-score equation}
		\begin{equation}
		U_{k}(\theta)=\sum_{i=1}^{n}\sum_{t=\tau_0+1}^{k} \big(  \frac{ \partial \tilde{\mu}_{it}^{(k)} }{\partial \theta} \big)^T \big(  \tilde{\nu}_{it}^{(k)}    \big)^{-1} (I_{it}- \tilde{\mu}_{it}^{(k)})
		\label{quasi-score equation measurement error}
		\end{equation}
		\textbf{using iterative methods, e.g. Newton-Raphson and iteratively reweighted least squares, where $\tilde{\mu}_{it}^{(k)} =\tilde{\nu}_{it}^{(k)} \triangleq \tilde{R}_{it}^{(k)} \Lambda_{it}$.}
		\State \textbf{Obtain $\hat\beta^{(k)}$ and $\{\hat R_{it}^{(k)}\}_{1\leq t\leq k}^{1\leq i\leq n}$ by plugging $\hat{\theta}^{(k)}$ into (\ref{betatilde}) and (\ref{tildeY}), i.e.,}
		\begin{align}
		\hat{\beta}^{(k)} &\triangleq \left\{ \sum_{i=1}^{n} \sum_{t=1}^{k-1} (Z_{it}-\bar{Z}_{k-1})(Z_{it} -\bar{Z}_{k-1})^T        \right\}^{-1} \label{estimatorforbeta}\\
		&\qquad \times \sum_{i=1}^{n} \sum_{t=1}^{(k-1)} (Z_{it} -\bar{Z}_{k-1}) \Big[ h(\hat{R}_{it}^{(k-1)}) - \sum_{m=1}^q \hat{\theta}_m^{(k)} f_m \big(D_{1,it},D_{2,it},\hat{D}_{3,it}^{(k-1)} \big) \Big],   \nn  \\
		{\text{ and }}	&\hat{R}_{it}^{(k)}  \triangleq h^{-1} \left( \sum_{m=1}^q \hat{\theta}_m^{(k)} f_m (D_{1,it},D_{2,it},\hat{D}_{3,it}^{(k)}) + Z_{it}^T \hat{\beta}^{(k)} \right).  \label{estimatorforY}
		\end{align}
		\EndFor
		\EndLoop
		\EndProcedure
		\Statex
	\end{algorithmic}
	\vspace{-0.4cm}%
\end{algorithm}

\subsection{Uncertainty quantification}
We use the block bootstrap method introduced by \cite{hall1985resampling} and \cite{kunsch1989jackknife} to quantify the uncertainty of the proposed online estimator. We will illustrate this calculation for $\hat\beta$. The confidence band of $\{ \hat{R}_{it} \}_{1\leq t\leq N}^{1\leq i\leq n}$ can be obtained similarly.  Given the time series data of daily case incidence and local area factors, denoted as$\{  (I_{it},Z_{it}) \}_{1\leq t\leq N, 1\leq i \leq n}$, where $Z_{it}$ is a $p$-dimensional vector of covariates at time $t$ in location $i$, we assume that $\beta$ can be estimated using a block of the time series data as long as the length of the block exceeds $\ell$. Specifically, we require that the expression
$\left( \sum_i \sum_{t\in \mathcal{S}_i}  (Z_{it}-\bar{Z}_{\mathcal{S}})(Z_{it} -\bar{Z}_{\mathcal{S}} )^T   \right)^{-1} $ be well defined for any arbitrary set $\mathcal{S}_i\subset \{1,\cdots,N\}$ with $|\mathcal{S}_i|\geq \ell$, $i=1,\cdots,n$ and $\bar{Z}_{\mathcal{S}}$ is defined as an analogy to $\bar{Z}$.

For $B$ bootstrapping samples, we assume the sampled block $\mathcal{S}_b$ is the same for all counties in one bootstrapping observation for simplicity, $b=1,\cdots, B$. That is, we denote a block of data up to time $t$ with length $\ell$ as 
\begin{equation*}
\xi_{t}=\Big\{ \big( I_{i,t-\ell+1},Z_{i,t-\ell+1}  \big), \cdots , \big(  I_{it},Z_{it}  \big), i=1,\cdots,n  \Big\}
\end{equation*}
for $t=\ell,\cdots,N$. We adopt the suggested $\ell=O(N^{1/3})$ from \cite{buhlmann1999block} and independently resample $B$ blocks of data with replacement, for $B=O( \lfloor (N-p-\tau_0)/\ell \rfloor)$ \citep{buhlmann2002bootstraps} to obtain $B$ bootstrap samples $\xi_{t_1}, \xi_{t_2},\ldots, \xi_{t_B}$. 

For the effect of $j$-th covariate, $\beta_j$, where $j=1,\cdots,p$, let's denote the estimator based on the $b$-th sample as $\hat{\beta}^*_{t_b,j}$ and the estimator using the whole data as $\hat{\beta_j}$. One may use $\{ \hat{\beta}^*_{t_b,j} - \hat{\beta}_j : b=1,\cdots,B \}$ to approximate the empirical distribution of $( \hat{\beta}_j - \beta_{j})$ or use $\{ \hat{\beta}^*_{t_b,j}: b=1,\cdots,B \}$ as an approximation of the empirical distribution of $\hat{\beta}_j$. Thus, the level $1-\alpha$ bootstrap confidence interval of $\beta_j$ based on the $B$ samples, denoted as $(L_{j,\alpha/2,B}^*, U_{j,\alpha/2,B}^*)$, can be constructed by
\begin{gather}
L_{j,\alpha/2,B}^*=2\hat{\beta}_j - \sup\left\{   t: \f{1}{B} \sum_{b=1}^B \mathbbm{1}(\hat{\beta}^*_{t_b,j} \leq t) \leq 1- \f{\alpha}{2}        \right\}, \nn \\
U_{j,\alpha/2,B}^*= 2\hat{\beta}_j - \inf\left\{   t: \f{1}{B} \sum_{b=1}^B \mathbbm{1}(\hat{\beta}^*_{t_b,j} \leq t) \geq  \f{\alpha}{2}        \right\}, \;\; \text{or}  \label{Boot1}
\\
L_{j,\alpha/2,B}^*=\inf\left\{   t: \f{1}{B} \sum_{b=1}^B \mathbbm{1}(\hat{\beta}^*_{t_b,j} \leq t) \geq  \f{\alpha}{2}        \right\} , \;
U_{j,\alpha/2,B}^*= \sup\left\{   t: \f{1}{B} \sum_{b=1}^B \mathbbm{1}(\hat{\beta}^*_{t_b,j} \leq t) \leq 1- \f{\alpha}{2}        \right\} \nn .
\end{gather}

\section{Properties of the Estimators and Algorithm} \label{properties} In this section, we show the asymptotic results of the estimators and property of the estimating procedure.

\subsection{Asymptotic properties}
The proposed method is a general approach for various counting processes and is robust to mis-specification of distribution assumptions. When the measurement error can be ignored, i.e., $\epsilon_{it}=0$, the Poisson process used in \citep{cori2013new} becomes a special case of Equation (\ref{general GLM setting}), with $\nu_{it}=\mu_{it}$. Since $U_t(\gamma)\in \mathcal{F}_{t} \triangleq \sigma( \cup_{i=1}^n \mathcal{F}_{i,t} )$ and $\partial\mu_{it}/\partial \gamma, \mu_{it},\nu_{it} \in \mathcal{F}_{i,t-1}$, then ${ (U_s(\gamma),\mathcal{F}_s), s\geq q }$ forms a mean zero martingale sequence, indicating that $U_N(\gamma)$ is an unbiased estimating function. For asymptotic properties of the estimators, we consider the least favorable case where we only have one county's data and discard subscript $i$ in all notations. Denote the true parameter value as $\gamma_0$, and according to Theorem 3 of \cite{kaufmann1987regression}, we have:

\begin{thm}
	Under condition 1 described in Section B of the Supplementary Material and on the non-extinction set defined in (B.1), 
	\begin{equation}
	\Big[ \sum_{t=q}^{N}\cov  \big(   \xi_{t}(\gamma_0) | \mathcal{F}_{t-1}    \big)	  \Big]^{1/2} (\hat{\gamma}-\gamma_0) \overset{d}{\longrightarrow} \mathcal{N}(0,I). \label{normal}
	\end{equation}
\end{thm}

\noindent In the Supplementary Materials, we also show a special case of (\ref{model_assumption}), given by
\begin{equation}\label{log_special_case}
\log (R_{it})= Z_{it}^T \beta + \theta_{q}+ \sum_{m=1}^{q-1} \theta_m \left[  \log (R_{i,t-m}) -Z_{i,t-m}^T \beta   \right], \; {\rm for}\; t > q >0, 
\end{equation}
where $\Vert \theta \Vert_1 <1$ and $\theta_m\geq 0$ for $m<q$ satisfies (\ref{normal}) under a more traceable condition {2}.

\subsection{Bias correction and property of the estimation procedure}
\label{model property}

When the measurement error cannot be ignored, i.e., $\epsilon_{it}\neq 0$, the proposed iterative algorithm consists of two major steps: one that utilizes a semi-parametric locally efficient estimator to express $\beta$ and $R_{it}$ as functions of $\theta$, and another that updates the estimates by solving the score equation. However, the latter step is subject to estimation bias due to the non-zero measurement error term in equation (\ref{model_assumption measurement error}), which can cause bias in the score equation \citep{cameron2013regression}. Correcting this bias in the general case represented by (\ref{model_assumption measurement error}) is challenging, but it can be tackled in the special case described by (\ref{non-stationary special case}) by requiring $\e \big( e^{\epsilon_t} |\mathcal{F}_{t-1} \big) = c_0$ for some constant $c_0$ \citep*{zeger1988regression}. Therefore, we can correct the bias by modifying the quasi-score equation (\ref{quasi-score equation measurement error}) into an unbiased estimating equation as follows:
\begin{equation*}
U_{k}^{*}( \theta)=\sum_{t=\tau_0+1}^{k} \big(  \frac{ \partial \tilde{\mu}_t^{(k)} }{\partial \theta} \big)^T \big(  \tilde{\nu}_t^{(k)}    \big)^{-1} (I_t/c_0- \tilde{\mu}_t^{(k)}). \label{adjusted quasi-score equation measurement error}
\end{equation*}
%which is unbiased since
%\begin{equation*}
%        \f{1}{c_0}\e \big( I_t |\mathcal{F}_{t-1} \big) - \tilde{\mu}_t^{(k)} \big\vert_{\theta,\beta} = \Lambda_{t} e^{  Z_t^T \beta +  \sum_{i=1}^q \theta_i   \log (R_{t-i})}   \Big(  \f{1}{c_0} \e \big( e^{\epsilon_t} |\mathcal{F}_{t-1} \big) -1 \Big)=0.
%\end{equation*}

In practice, $c_0$ can be treated as a nuisance parameter and estimated by solving the quasi-score equation. Moreover, under Equation (\ref{non-stationary special case}), solving the modified quasi-score Equation (\ref{quasi-score equation measurement error}) is equivalent to obtaining
\begin{equation}
\hat{\theta}^{(k)}=\arg\max_{\Vert \theta \Vert_1 <1} \sum_{t=\tau_0+1}^{k} \Big( ( I_t / c_0 ) \log \tilde{R}_t^{(k)}  - \tilde{R}_t^{(k)} \Lambda_{t} \Big),  \label{estimator for theta}
\end{equation}
where $\hat{\theta}^{(k)}$ is the MLE based on conditional profile likelihood of Poisson counts. Therefore, the proposed online estimation procedure guarantees the following properties. The proof of the theorems is presented in Section C of the Supplementary Materials.

\begin{thm}[Concavity] When $k> \tau_0$, updating the estimates in each iteration, which is equivalently to solve equation (\ref{estimator for theta}), is a globally concave maximization problem.
	\label{thmconcave}
\end{thm}

\begin{thm}[Iteration Difference] \label{updatesbound}
	Given regularity condition {3} in Section C of the Supplementary Materials, using data of observed case incidence and local-area factors up to time $N$, $\{ (I_{t}, Z_{t} ) \}_{1\leq t \leq N}$, for each $1\leq m\leq q-1$,
	\begin{equation*}
	\big| \hat{\theta}_{m}^{(k)}- \hat{\theta}_m^{(k-1)} \big| = O\Big(   \f{1 }{k-1}+\f{I_{k}}{\big( \sum_{t=\tau_0+1}^{k-1}  I_{t} \big)}     \Big),
	\end{equation*}
	where $k \leq N$ is the indicator of iteration.
\end{thm}

%\begin{remark}
Theorem \ref{updatesbound} shows that the bound of the step-wise difference of the online estimator between iterations decreases as the number of observation time increases. The consistency of the estimator remains an open question as discussed in \cite{davis2003observation}. The difficulty arises from the measurement error, $\epsilon_{it}$, and the series dependency of disease incidence, $\Lambda_{it}$, in the model. Both elements are essential in modeling the dynamics of disease transmission. Similar but simplified models, which do not model the measurement errors and series dependency of disease incidence were studied for different contexts, e.g. studies of economics or disease without series dependency of disease incidence \citep{davis1999modeling,davis2000autocorrelation, davis2003observation,fokianos2009poisson,neumann2011absolute,doukhan2012weaka,doukhan2012weakb}. Inspired by the derivation and discussion in \cite{davis2003observation}, we conjecture that conditions requiring stationary and uniformly ergodic of $\log(R_{t}\Lambda_{t})$ are needed to establish the consistency of estimators of model (\ref{non-stationary special case}), although these conditions may be difficult to verify in practice. 

\section{Simulation Studies}
\label{simulation}
This section describes the simulation studies we conducted to assess the performance of the proposed method. We focused on the estimation of the disease transmission rate and the impacts of time-varying factors . We calculated the relative bias, the coefficient of variation of the estimates, and the Bootstrap confidence interval coverage probability under various scenarios. Additionally, we compared the performance of the proposed model and the basic SIR model, the details of which are presented in Section A of the Supplementary Material.

\subsection{Simulation settings}
We generated data on the daily numbers of SARS-CoV-2 infections, along with associated local-area factors influencing the disease transmission rate. We generated data in a single location with time-varying factors and suppressed the location subscript $i$ in this section. We fit the proposed model, as described by Equation (\ref{non-stationary special case}), assuming that the $R_{t}$ values exhibited an AR(1) dependency over time and were linked with an intercept term and two time-varying factors, i.e. $q=2$ and $p=2$. Each scenario was replicated 1,000 times to calculate evaluation statistics. The infectiousness function was assumed to be known and was modeled using the probability density function of a gamma distribution, in a manner mirroring the results of a previous epidemiological study of SARS-CoV-2 cases in Wuhan, China \citep{li2020early}. The estimation algorithm was applied with a pre-estimation period of 5 days, i.e., $\tau_0=5$.

\begin{table}[h!]
	\caption{Parameter values used to generate data in simulation scenarios. In all scenarios, we set $\beta_{1}=-0.02$ and $\beta_{2}=-0.125$. Except for scenario 6, the standard deviation of the error terms in each model is determined by $(\var \epsilon)^{1/2} \asymp \e |Z_{T1}|\beta_{1}$. $Z_{t1}$ was generated from $4.5+(t-\f{T}{2})/6 +\mathcal{N}(0,9)$ in scenario 1 and generated from $9+(t-\f{T}{2})/16 +\mathcal{N}(0,9)$ in scenario 3 to avoid unrealistically large values of $R_t$.}
	\label{Parameter}
	\begin{center}
		\noindent 
		\vskip -1em
		\footnotesize{         
			\begin{tabular}{c l l l l l l l}\hline 
					Scenario 1: &$T=90,$  & $ I_0=500,$ & $ \epsilon\sim N(0,10^{-3}),$  &$\theta_2=0.5,$ & $\theta_1=0.7,$  &   \\ \hline
				Scenario 2: & $T=120,$  & $ I_0=500,$ & $ \epsilon\sim N(0,10^{-3}),$ &$\theta_2=0.5,$ & $\theta_1=0.7,$ &   \\ \hline
				Scenario 3: &$ T=200,$ & $I_0=500,$ & $\epsilon\sim N(0,10^{-3}),$ &$\theta_2=0.5,$ & $\theta_1=0.7,$  &   \\ \hline
				Scenario 4: & $ T=120,$ & $I_0=125,$ & $ \epsilon\sim N(0,10^{-3}),$  &$\theta_2=0.5,$ & $\theta_1=0.7,$  & \\ \hline
				Scenario 5: & $ T=120,$ & $I_0=250,$ & $ \epsilon\sim N(0,10^{-3}),$ &$\theta_2=0.5,$ & $ \theta_1=0.7,$  & \\ \hline
				Scenario 6: & $ T=120,$ & $ I_0=500,$ & $ \epsilon\sim N(0,10^{-2}),$ &$\theta_2=0.5,$ & $ \theta_1=0.7,$  &  \\ \hline
				Scenario 7: &$ T=120,$ & $ I_0=500,$ & $ \epsilon\sim t_3/10^{3/2},$ &$\theta_2=0.5,$ & $\theta_1=0.7,$  &   \\ \hline
				Scenario 8: &$T=120,$ & $ I_0=500,$ & $\epsilon\sim N(0,10^{-3}),$  &$\theta_2=0.5,$ & $\theta_1=0.7,$ & $ \beta_3=-0.03,$     \\ \hline
				Scenario 9: & $T=120,$ & $ I_0=500,$ & $ \epsilon\sim N(0,10^{-3}),$  &$\theta_3=0.45,$ & $ \theta_1=0.5,$ & $ \theta_2=0.3,$    \\ \hline
			\end{tabular} 
		}
	\end{center}
\end{table}

Data were simulated under various scenarios, including settings where the fitted models were correctly specified or misspecified, as shown in Table \ref{Parameter}. For the scenarios with correctly specified models (scenarios 1-7), data were generated using the parameters $(\theta_1,\theta_2, \beta_1,\beta_2)=(.7,.5, -.02,-.125)$. Here, the simulated values $\{ R_{t}, t\geq 1\}$ were designed to mimic the trend of the $R_{it}$ reported in \cite{pan2020association}. We varied factors such as the days of observation ($T$), initial incident cases ($I_0$), and utilized different distributions of error terms ${ \epsilon_t\sim_{i.i.d}\epsilon, t=1,\ldots, T}$ for data generation. In the settings where the model was mis-specified (scenarios 8-9), data were simulated with the assumption that the $R_t$ values followed an AR(2) dependency (i.e., $q=3$) or were associated with an intercept term and three covariates (i.e., $p=3$). The intercept term is $\theta_q$, while the two time-varying covariates ($Z_{t,1},Z_{t,2}$) were simulated independently to mimic the real data of temperature in Philadelphia and data of social distancing from daily cellular telephone movement, provided by Unacast \citep{unacast}, which measured the percent change in visits to nonessential businesses, e.g., restaurants and hair salons during March 1st and June 30th, 2020. Specifically, daily temperatures were generated from a shifted normal distribution, i.e., $Z_{t1}\overset{d}{=} 5+ \Big(t-\f{T}{2}\Big) \big/ 8 +\mathcal{N}(0,9)$, for $ t=1,\cdots,T$. The social distancing data were generated from a uniform distribution and applied a logit transformation afterwards, i.e., $Z_{t2}\overset{i.i.d}{\sim } 2+logit(Uniform(0,1)),$ for $ t=1,\cdots,T$. Additionally, in the misspecified scenarios with three covariates, the third covariate was generated from $Z_{t3}\overset{d}{=} \Big(t-\f{T}{2}\Big) \big/9 +\mathcal{N}(0,5)$, for $ t=1,\cdots,T$. The variations of error terms were assumed to be small as the variation of $\log(R)$ was small with $R$ ranging from 0.5 to 3 and has a comparable size of the variation of covariates times their effects.

\begin{table}[ht]
	\centering
	\caption{Bias and coefficient of variation for the estimators in each scenario. The relative bias, bias, and coefficient of variation of the estimators were calculated using the following formulas. Relative bias:$\sum_{i=1}^{N_s}(\hat{a}_i- a)/a N_s$. Bias: $\sum_{i=1}^{N_s}(\hat{a}_i-a)/N_s$,Coefficient of variation: $\{ \sum_{i=1}^{N_s}[\hat{a}_i - (\sum_{i=1}^{N_s} \hat{a}_i /N_s ) ]^2 \}^{1/2}  /  (\sum_{i=1}^{N_s} \hat{a}_i /N_s ) $. Here $N_s=1000$ represents the number of replicates, $a$ denotes the true value of a parameter, and $\hat a$ indicates the estimated value of that parameter.}
	\label{Estimation results}
	{\tabcolsep=4.25pt
		%\begin{tabular}{|c|c|c|c|c|c|c|c|c|c|c|c|c|}
		\begin{tabular}{@{}ccccccccccccccc@{}}
			\hline
				&\multicolumn{4}{c}{relative bias (\%)}  && \multicolumn{4}{l}{empirical bias ($\times 10^{-2}$)} & &\multicolumn{4}{l}{coefficient of variation $\hat{\sigma}/\hat{\mu}$} \\ 
				\hline
				\bf{Scenario}& $\theta_1$ & $\theta_{2}$
				&$\beta_1$ &$\beta_2$  && $\theta_1$ & $\theta_{2}$
				&$\beta_1$  &  $\beta_2$  && $\theta_1$ & $\theta_{2}$
				&$\beta_1$ &  $\beta_2$ \\ 
							\bf{1} &-1.51  &-2.94  & -1.19& 6.26 && -1.05 &-1.47 & -0.02& 0.78 &&0.10 &0.37 &0.55&0.38 \\ 
			\bf{2} &-1.42 &-2.41  & -0.34& 5.03 && -0.99 &-1.20 & -0.01& 0.63 &&0.11 &0.27 &0.36&0.36 \\
			\bf{3} &-0.78  &-1.15 & -0.47& 2.19 && -0.55 &-0.57 & -0.01& 0.27 &&0.10 &0.24 &0.27&0.29 \\ 
			\bf{4} &-3.57 &-2.08  & -0.04& 7.18 && -2.50 &-1.04 & -0.00& 0.90 &&0.14 &0.37 &0.69&0.64 \\ 
			\bf{5} &-1.85 &-4.87  & -5.67& 5.00 && -1.29 &-2.44 & -0.11& 0.63 &&0.13 &0.39 &0.98&0.60 \\ 
			\bf{6}&-4.16  &-1.10  & -4.47& 6.90 && -2.91 &-0.55 & -0.09& 0.86 &&0.13 &0.39 &0.43&0.48 \\ 
			\bf{7} &-1.43 &-4.20  & -0.53& 8.22 && -1.00 &-2.10 & -0.01& 1.03 &&0.11 &0.28 &0.37&0.36 \\ 
			\bf{8}  & -0.72 &-17.83 & 4.06  & 9.64 &&-0.51 &-8.92 & 0.14& 1.20 && 0.12 & 0.31 & 0.31& 0.42 \\ 
			\bf{9}&$\diagup$ & $\diagup$ & -17.67& 9.95 && $\diagup$ & $\diagup$& -3.53 & 1.24  && $\diagup$& $\diagup$& 0.27& 0.27\\
			\hline
		\end{tabular} 
	}
\end{table}

\subsection{Simulation results}

%enrich the table caption. Can we summarise simulations to three categories: 1. explore n and starting I, 2. explore varing sigma (justify the use of small sigma) by calculating the R variation without sigma, if R is uniform betwen 0.7 to 2, the variation is about 0.08, 3. explore measurement errors. For relative bias and actual bias, can be report one? } 

Table \ref{Estimation results} presents the bias and coefficient of variation of the estimators for each scenario. In scenarios where the models were correctly specified, both the bias and the coefficient of variation decreased with an increase in either the days of observation (scenarios 1-3) or the initial case count (scenarios 4-5). When the variance of the error term increased to a magnitude smaller than the product of the time-varying local-area factors and their corresponding effect sizes (comparing scenarios 6 and 2), the proposed estimator still demonstrated robust performance, with less than a 5\% increase in relative bias and a 0.2 increase in the coefficient of variation. Even when the error term exhibited a heavier tail (comparing scenarios 7 and 2), the relative bias increased by less than 4\%, and the coefficient of variation increased by less than 0.01.

\begin{figure}[!ht]
	\centerline{\includegraphics[width=6in]{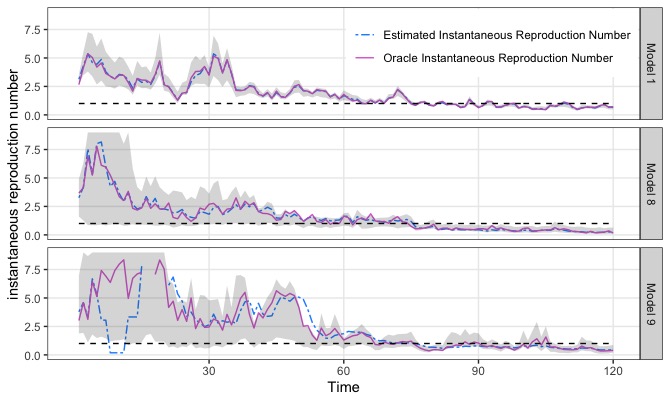}}
	\caption{Estimated $R_t$'s and 90\% bootstrap confidence bands from random replications of simulation, with block length $\ell=45$ and bootstrap replication $B=200$.}
	\label{ConfidenceBand}
\end{figure}

Two types of mis-specification were considered. In scenario 8, an important time-varying covariate was omitted from the model. Due to the correlation between the generated local-area factors over time, this scenario differs from merely increasing the error term's variance. In both mis-specified cases, the estimator $\hat{\theta}$ is closer to the corresponding partial correlation coefficients rather than to the true values. Although the mis-specification led to an increase in estimation bias, the algorithm still provided a relatively accurate estimator for the correctly specified parts, showing a small bias and coefficient of variation (comparing scenarios 2, 8, and 9). Most importantly, as illustrated in Figure \ref{ConfidenceBand} from a random replication of the simulation, the estimated $R_{t}$ values still captured the tendency of the true $\{ R_{t} \}_{t\geq 0}$, especially when the time-series sample size increased.

Two types of mis-specification were considered. In scenario 8, an important time-varying covariate was omitted from the model. Owing to the correlation between the generated local-area factors over time, this scenario represents more than just an increase in the error term's variance. In both cases of mis-specification, the estimator $\hat{\theta}$ was closer to the corresponding partial correlation coefficients than to the true values. While the mis-specification led to increased estimation bias, the algorithm still provided a relatively accurate estimator for the correctly specified parts, exhibiting a small bias and coefficient of variation (as seen when comparing scenarios 2, 8, and 9). Most importantly, as illustrated in Figure \ref{ConfidenceBand} from a random replication of the simulation, the estimated $R_t$ values still accurately captured the trend of the true $\{ R_t \}_{t\geq 0}$, particularly as the time-series sample size increased.

\begin{table}[h!]
	\caption{Coverage Probability of $1-\alpha$ Empirical Bootstrap Confidence Interval for $\theta_2$.}
	\label{Bootstrap inspection performance}
	\begin{center}
		{\tabcolsep=4.25pt
			%\begin{tabular}{|c|c|c|c|c|c|c|c|c|c|c|c|c|}
			\begin{tabular}{@{}ccccccc@{}}
				\hline
					&  & T=300 & T=300  & T=300 & T=450 & T=450  \\
					& & {$\ell=30$} & {$\ell=45$}  & {$\ell=60$}  &  {$\ell=30$}  & {$\ell=45$}  \\
					\hline
				&$\alpha=5\%$ & 96.4\%   & 94.8\% & 93.0\% & 99.8\% & 99.2\%  \\
				&$\alpha=10\%$ & 91.2\%  & 87.8\% & 86.2\% & 98.6\%  &  97.2\% \\
				& $\alpha=20\%$ & 83.4\%   & 79.8\% & 77.2\%  & 96.8\%  & 93.2\% \\
				\hline
			\end{tabular}
		}
	\end{center}
\end{table}

We also constructed the bootstrap confidence interval according to (\ref{Boot1}), and the coverage probability of the bootstrap confidence interval for $\theta_2$ is shown in Table \ref{Bootstrap inspection performance} as an illustration. Table \ref{Bootstrap inspection performance} confirms that the proposed bootstrap confidence interval provides the expected coverage probability when choosing $\ell=45=\tau_0+O(T^{1/3})$ for $T=300$, following the recommendations in \cite{buhlmann1999block}, and setting the number of replications to $B=200=O(T^{2/3})$, as per \cite{buhlmann2002bootstraps}. When a smaller block length $\ell=30$ is chosen for $T=300$, the accuracy of the bootstrap estimator is reduced, leading to a conservative confidence interval. In contrast, when a larger block length $\ell=60$ is chosen, the bootstrap estimator's accuracy may improve, but the coverage probability becomes less than ideal due to the high correlation between bootstrap subsamples. Furthermore, the optimal block length is also related to $T$, as $\ell=30$ and $\ell=45$ would only result in conservative confidence intervals for $T=450$. Practically, we suggest choosing the smaller value of $\ell$ that satisfies $\ell=O(N^{1/3})$ as recommended in \cite{buhlmann1999block}.

\section{Application to COVID-19 Data}
\label{real data}
We utilized the proposed method to analyze a dataset that includes data from the early stages of the COVID-19 pandemic. The dataset contains daily infection data and county-level risk factors from 808 US counties across 47 states and the District of Columbia, spanning from March 2020 to May 2021. We examined the impact of county demographics, social behavior, and vaccination coverage on disease transmission and provided projections of daily COVID-19 case counts for three distinct pandemic waves, including the first one. For each period, the model was trained using observed daily infection and covariate data for 4 months (training window) and made projections for the subsequent four weeks (projection window). The first two periods, from March to July 2020 and August to December 2020, covered the first wave and the first holiday season of the pandemic, respectively. During these periods, we used county-level population density, daily wet-bulb temperature, and social distancing measures to explain the variation of $R_{it}$ and project future cases. These decisions were informed by studies identifying factors that affect disease transmission and the predominant use of social distancing measures at the time \citep{rubin2020association, talic2021effectiveness, sera2021cross, hou2021intracounty, weaver2022environmental}. The third period, from January to May 2021, coincided with when COVID-19 vaccines became publicly available. In this phase, we incorporated vaccination coverage data along with the three existing covariates to model variation of $R_{it}$ and project future cases.

\begin{figure}
	\caption{Locations of the counties included in our analysis. Counties were categorized as small, midsize, and large using population cutoffs of 100,000 and 250,000 people.}
		\label{map}
	\begin{center}
		\includegraphics[width=1\linewidth]{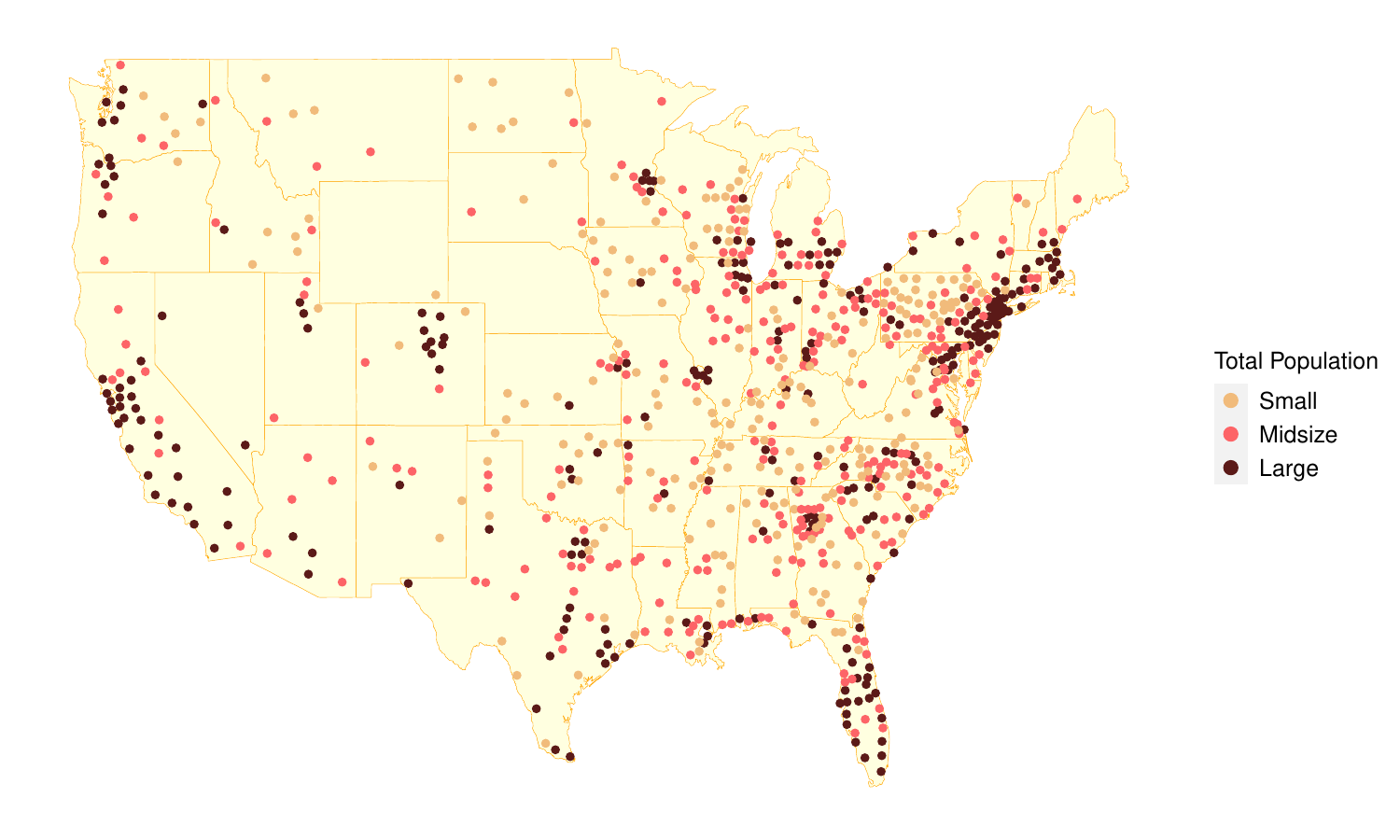}
	\end{center}
\end{figure}

To be included in this analysis, a county had to have 5 incident cases for more than two out of six consecutive days for at least 20 times, as of June 1st, 2020 and meet at least one of the following criteria: 1)contain a city with a population exceeding 100,000; 2)contain a state capital; 3)be the most populated county in the state or 4) have an average daily case incidence exceeding 10 during June 1st to June 30th, 2020. Counties were also required to have social distancing data for each of the study period and vaccination data from the CDC for the third study period. The 808 counties represent 79.5\% of the US population, as shown in Figure \ref{map}. Population density data was sourced from the US Census data and is expressed as number of people per square mile. Log transformation was performed followed by standardization due to the large values of this variable and substantial skewness in density for the largest cities. Social distancing, obtained from Unacast, was measured by the percent change in visits to nonessential businesses (e.g., restaurants, hair salons), revealed by daily cell phone movement within each county, compared with visits in a four-week baseline period between February 10th and March 8th, 2020. We employed a rolling average of the percentage of visits from 7 to 14 days prior to the time of interest to account for the potential lag between changes in social distancing and alterations in disease transmission. Vaccine coverage data, obtained from the CDC, was defined as the percentage of the population that has received at least one vaccine dose. We applied a 14-day lag to allow time for any potential effects of the vaccine to manifest. The percentage of the non-vaccinated population, calculated by subtracting the 14-day-lagged vaccination percentage from 100\%, was then entered into the regression model. When projecting future case numbers, we used county population density, historical average wet-bulb temperature from the last 10 years before the analysis year, and the most recent weekly average values for social distancing and vaccination coverage to calculate the assumed covariate values for future days.

%Within each county, the cumulative case count before the first $\varrho$ incidence cases was considered as $I_0$. As a sensitivity analysis, we explored three different $\varrho$ values, i.e., $\varrho=5,10,15$, and since the analyses showed similar results, we presented the results with $\varrho=10$. The incidence threshold of 20 in criterion 4) was selected based on the empirical distribution of case counts in the observation time period. Daily incidence case counts were smoothed using a 4-day moving average to reduce the impact of batch reporting. Our goals were to examine the association between county-level factors and $R_{it}$, to forecast the future case count, and to provide evidence for selecting appropriate social distancing policies.

%\subsection{Covariate selection and data preparation}

\subsection{Influential local-area covariates for disease transmission}

The estimated covariate effects on $R_{it}$ with bootstrap intervals for the three time periods are shown in Table \ref{Estimates}. Among the four local-area covariates we examined, social distancing appeared to be the most influential factor related to disease transmission, particularly during the first wave of the pandemic. It exhibited a significant positive association with disease transmission. During the first wave, a 50\% reduction in the frequency of visiting non-essential businesses was estimated to reduce $R_{it}$ by an average of 11\%. This finding aligns with the effects of social distancing reported in other studies from the first wave of the pandemic, but our study encompasses more counties across the US \citep{rubin2020association,courtemanche2020strong}. Over time, the impact of social distancing diminished. This change could be attributed to multiple factors, including possible changes in the virus and variations in population behavior, which might have prevented our social distancing measure from capturing all aspects of population mobility. Furthermore, we observed a gradual decrease in social distancing over time, from a median of 31\%  reduction in the first wave to 11\%  reduction in the third wave. It's also possible that the effect of social distancing is nonlinear, and its impact was minimal when only a small amount of social distancing was implemented. 

We also observed a significant heterogeneity in disease transmission across counties with varying population densities during the first wave. The reproduction number was higher in more populated areas compared to rural areas. Simultaneously, we found that wet-bulb temperature was negatively associated with $R_{it}$, suggesting that transmission was generally higher in colder weather compared to warmer days. Although the regression coefficient of temperature was small, its impact on policy-making could be substantial due to the wide range of temperatures a county might experience within a year. During the first study period, the largest increment in wet-bulb temperatures observed within a county was 21$^{\circ}$C, which could reduce COVID-19 transmission by 15\%, with other covariates held constant. However, temperature could also affect social distancing values, as outdoor activities tend to increase during spring and summer, potentially offsetting the reduction in transmission due to increased temperatures. Both the heterogeneity associated with varying population densities and the impact of temperature decreased over time, indicating that the entire US began to experience a more uniform pandemic situation with less seasonality as we moved into the middle or later stages of the pandemic. Additionally, the effect of vaccination coverage on disease transmission during the first half of 2021 was not found to be significant. This lack of significance could be due to factors such as relatively lower vaccine coverage during early-stage vaccination rollout,  behavioral changes post-vaccination, and challenges in achieving herd immunity.

\begin{table}[htbp!]
	\newcolumntype{P}[1]{>{\centering\arraybackslash}p{#1}}
			\caption{Model estimates and projections for the COVID-19 data during three study periods. Bootstrap intervals were based on replication $k=200$ and block length $\ell=45$.}
				\label{Estimates}
	\begin{tabular}{lP{30mm}P{30mm}P{30mm}}
		{\textbf{study period}} & {\textbf{1}} & {\textbf{2}} & {\textbf{3}}\\
		{\textbf{training window}} & 03/01/2020-06/30/2020 &08/01/2020-11/30/2020 & 01/01/2021-05/31/2021 \\
		{\textbf{projection window}}&07/01/2020-07/28/2020  &12/01/2020-12/28/2020& 06/01/2021-06/28/2021\\
		{\textbf{covariates}}& population density \newline temperature \newline social distancing &population density \newline temperature \newline social distancing &population density \newline temperature \newline social distancing \newline vaccination coverage\\
		{\textbf{covariates effects}} &&&\\
		population density($\times 10^{-2}$)  &7.32 (1.89, 3.19) &2.37 (1.14, 4.74) &0.29 (-0.04, 0.59)\\
		temperature ($\times 10^{-2}$)&-0.76 (-1.69, -0.16) &-0.36 (-0.83, -0.09) & -0.16 (-0.82, 0.05)\\
		social distancing &0.24 (0.10, 0.85)&0.15 (0.07, 0.24) & 0.03 (-0.31, 0.15)\\
		proportion of  non-vaccinated &--- &---& 0.02 (-0.20, 0.03)\\
		{\textbf{mean PAE (\%)}} &&&\\
		week 1 &27.9&20.8&24.1\\
		week 2 &25.5&20.2&27.4\\
		week 3 &22.5&23.9&34.2\\
		week 4 &24.6&40.5&39.1\\
		{\textbf{median  PAE (\%)}} &&&\\
		week 1 &22.5&17.8& 16.4\\
		week 2 &19.4&16.6&17.9\\
		week 3 &17.7&16.7&19.9\\
		week 4 &20.1&27.3&25.1\\
	\end{tabular}
\end{table}

\subsection{Forecasting future transmission}

%\begin{figure}[htbp!]
%	\centering
%	\includegraphics[width=1\linewidth]{pics//LargeRandPrediction.jpg}
%	\caption{{\small Heat maps for estimated and predicted county-level $R_{it}$ across United States. Roughly, $0< R_{it} <0.9$ : bluish, $0.9\leq R_{it} \leq 1.1$ : yellowish, $R_{it}>1.1$ : Reddish. Column A)  shows estimated $R_{it}$ at three time points, which illustrates the shift of epicenter during April to June 2020. Column B) demonstrates predicted $R_{it}$ on September 1st based on three levels of social distancing}}
%	\label{Prediction}
%\end{figure}

We compared the projected daily case counts with the actual observed numbers for the three study periods, and the accuracy of the projection was assessed using the weekly Percentage Absolute Error (PAE). Mean and median PAE are shown in Table \ref{Estimates} by week and study period. The mean PAE ranges from 20.2\% to 27.9\% for the first two weeks, and  22.5\% to 40.5\% for the second two weeks, respectively. We also compared our projection results with the CDC COVID-19 Cases Ensemble Forecast, which combines forecasts submitted by a large and variable number of contributing teams using different modeling techniques and data sources \citep{ray2020ensemble,cramer2022united}. During the three study periods, our model showed similar performance to the CDC model in the 1-week projection and was superior in the 3-4 week projections. The CDC model's average PAE across all states and weeks was 22\%, 32\%, 44\%, and 57\% for the 1 to 4-week forecast windows, respectively \citep{du2022deep}. It's worth noting that the CDC ensemble model is designed for state-level projection while our model focuses on local-level, specifically county-level, projections. Projecting at the county level is more challenging due to the smaller case incidence numbers and higher noise-to-signal ratio in county-level data.

\section{Discussion}
\label{discussion}
In this paper, we introduced a regression model with a measurement error term to monitor disease transmission and evaluate the impact of local-area factors in the early phases of outbreaks when data quantity and quality are limited. The time series dependency among the incidence case counts was decomposed into an epidemiological component, captured by the TSI model, and a statistical component, characterized using a time-series regression model. We employed the quasi-score method, complemented by a measurement error term, to tackle challenges related to poor data quality and potential model mispecification. We developed the iterative LOCAL-QUEST method to calculate online estimators of disease transmission and the impact of local-area factors on it. 

Our method provides another novel tool to a much larger toolbox for combating future outbreaks, particularly useful during the early stages or for small regional-level analysis when supplemental data resources are scarce. However, the challenges of modeling transmission during outbreaks are still not fully addressed by existing methods. While the measurement error term in our model partially accounts for the data errors and reduces bias, as demonstrated in our simulation analysis, our disease transmission estimates might still be biased. When additional data, such as seroprevalence and symptom onset data, become available, they should be incorporated to further correct biases caused by underreporting or reporting delays \citep{lison2023generative, quick2021regression}. Furthermore, identifying causal risk factors for disease transmission using our method remains challenging, necessitating specialized study designs for accurate inference.

While acknowledging the challenges and limitations of our approach, the proposed method offers flexibility for modifications and further extensions. For instance, one can introduce a lag time between the instantaneous reproduction number and local-area factors, e.g., by regressing $R_{it}$ on $X_{i(t-\tau_t)}$ with $\tau_t$ being a certain number of days in the regression equation, to account for delayed effects between changes in local-area factors and mean $R_{it}$ estimates across locations. It is also possible to incorporate a time series of under-reporting rates into the TSI model to adjust for varying under-reporting rates over time. When the under-reporting rate remains constant over time, using the reported cases counts to estimate $R_{it}$ is still valid \citep{cori2013new} as the constant under-reporting rate is canceled out in the TSI renewal process. Furthermore, the model can be extended into a nonparametric model by relaxing assumptions on the link function to avoid mis-specification. Further details are provided in Supplementary Material {A.6}. The estimation is straightforward if the dependency of $\{ R_{it}\}_{t\geq 0} $ is ignored for each $i$. When time series dependency of $\{ R_{it}\}_{t\geq 0} $ is considered, e.g., $\e \big[ R_{it} | Z_{it} , D_{3,it} \big] =g(Z_{it})+\sum_{m=1}^q \theta_m f_m (R_{i,t-m}) $ with a known $R_{i0}$ and known functions $f_m$, $1\leq m\leq q$ and unknown $g(\cdot)$, we can adopt local linear kernel estimators \citep{fan1996local} to estimate the parameters in the LOCAL-QUEST algorithm.

\bibliographystyle{imsart-nameyear} % Style BST file

\bibliography{RobustCovid}

\end{document}

\begin{supplement}
		\stitle{}
		\begin{description}
			
			\item Supplement to ``Early-Phase Local-Area Model for Pandemics Using Limited Data: A SARS-CoV-2 Application". 
			
			\item[R-code for Quasi-Score approach:] R-code together with part of the U.S. data used to analyze SARS-Cov-2 are available at https://github.com/Jiasheng-Shi/Covid-Quasi-Score. The file also contains R-codes for Bootstrap inspection for the instantaneous reproduction number estimation uncertainty.
			
		\end{description}

\section*{A. Extensions and additional results of simulation studies}

	\setcounter{equation}{0}
	\renewcommand{\theequation}{A.\arabic{equation}}
	
\subsection*{A.1. An extension to the setting of unknown variance function $g(\cdot)$}

Local polynomial fitting is one of the most commonly used procedures for tackling the issue of unknown variance function in quasi-likelihood models \citep{chiou1998quasi}. In the proposed model, we can also use this technique to approximate the variance function $g(\cdot)$ when it is unknown. Specifically, for a given kernel function $K(\cdot)$ and a bandwidth $h$, local polynomial fitting estimator of the $g(\cdot)$ is given by $$\hat{g}(\mu)=\hat{a}_{0}(\mu),$$ where $\hat{a}_0(\mu)$ and $\hat{a}_1(x)$ minimize the weighted square-loss
\begin{equation}
Loss\left(a_0,a_1 ; \mu, \{ \hat{\mu}_{it} \}_{1\leq t\leq N}^{1\leq i\leq n} \right) \triangleq \sum_{i=1}^{n} \sum_{ t= 1 }^{  \tau_0 + k }  \bigg[ \big(I_{it}- \hat{\mu}_{it}  \big)^2  -  a_1 \big(\hat{\mu}_{it} - \mu \big) -a_0  \bigg]^2 K \Big(  \f{ \hat{\mu}_{it} - \mu }{h}  \Big), \tag{A.1} \label{general_g}
\end{equation}
%and $\hat{\mu}_{it}$ is defined the same way as $\mu_{it}$ with parameter $(\beta,\theta)$ replaced by a given pair $(\hat{\beta},\hat{\theta})$. 
%The estimator can obtained by plugin this $\nu=\hat{g}(\mu)=\hat{a}_0(\mu)$ into the quasi-score equation.

\noindent where $\tau_0$ and $k$ are defined in the main manuscript Algorithm 1. Computing the estimates for this local polynomial fitting estimator proves to be somewhat time-consuming in our context. It is important to note that the loss function is defined with the provided values of $(\hat{\beta},\hat{\theta})$. Therefore, when attempting to solve the quasi-score equation (2.5) with an unknown variance function, we have an implicit form on the right-hand side of (2.5). For each specific $\theta$, we can compute $U_k(\theta)$ since the form of $g(\cdot)$ can be obtained when $\theta$ is known, and then search for the zero points. However, Newton-Raphson's method cannot be directly applied in this case.

Fortunately, applying such a local polynomial fitting estimator may not be essential in practice. For many instances where quasi-score estimation involves regression and covariates, as stated in \cite{liang1994use}, ``the quasi-likelihood method appears to be insensitive to the variance specification as the $\beta$ estimates and especially the corresponding robust standard error estimates are remarkably stable". 

A small simulation we conducted using observations from a single county is depicted in Figure.\ref{differ_g}, which supports the claim made by \cite{liang1994use}. The oracle instantaneous reproduction number is generated using $\nu=g(\mu)=\sqrt{\mu}$, while the estimated instantaneous reproduction number is calculated using the identity function as the variance function $g(\cdot)$. The simulation results show that the overall estimation of the instantaneous reproduction number is accurate, especially as the sample size grows. Even in the early time period, where the estimation and the oracle $R_{it}$ differ slightly, the estimates still capture the trend of the oracle $R_{it}$. This suggests the insensitivity of variance function in estimating the instantaneous reproduction numbers when regression structure and covariates information are involved, as is the case in our study.

\begin{figure}[h!]
	\renewcommand\thefigure{A.1}
	\centering
	\includegraphics[width=0.87\linewidth]{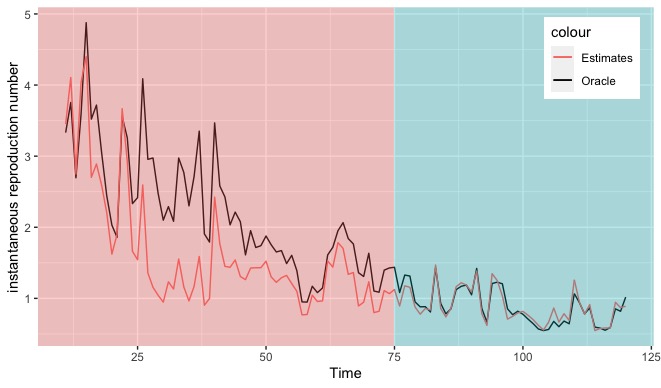}
	\caption{The oracle instantaneous reproduction number is generated using $\nu=g(\mu)=\sqrt{\mu}$, while the estimated instantaneous reproduction number is obtained using the identity function as the variance function. The results suggest that the estimation of the $R_{it}$ is largely insensitive to the specification of the variance function. Even in the early time periods, where the estimation slightly differs from the oracle $R_{it}$, the estimates still capture the trend of the oracle $R_{it}$.}
	\label{differ_g}
\end{figure}

\subsection*{A.2. Robustness comparison between the proposed model and a basic compartmental model}

In general, while classic epidemiological compartmental models can capture the dynamics of disease spread, as noted in \cite{quick2021regression} (and I quote), they ``are difficult to extend to flexible regression models on covariates." Hence, even though compartment models, like the classical SIR model, take suspected and recovered case data into account, they can not incorporate random structures like covariate data with measurement error. This omission potentially makes them less robust than the TSI model, as shown in our simulation, where we used observations from only one county and discarded subscript $i$ in all notations.

First, we tested our proposed LOCAL-QUEST method on the data generated from the SIR model in the following steps:
\begin{enumerate}[Step 1.]
	\item Generate the time-varying factors according to the simulation section of the main manuscript.
	\item Generate the oracle time series of the instantaneous reproduction number using (3.2).
	\item Generate recovery rate $\gamma_t$ from normal distribution $N(0.07,0.0025)$.
	\item Calculate the transmission rate $\beta_t=\gamma_t \cdot R_t$ and incident cases with a fixed initial incident case number $I_0$ and initial recovery case number $Rec_0$ through SIR model equations \citep{chen2020time}.
	\item Run LOCAL-QUEST algorithm with the generated time-varying factors and incident cases.
\end{enumerate} 
As shown in the upper panel of Figure.{A.2}, despite having a rough start on the estimation of the instantaneous reproduction number at the beginning, estimates based on our proposed LOCAL-QUEST method quickly captured the trend of $R_t$ and presented a rather robust performance after sample size increased over 60. 

\begin{figure}[h!]
	\renewcommand\thefigure{A.2}
	\centering
	\includegraphics[width=0.87\linewidth]{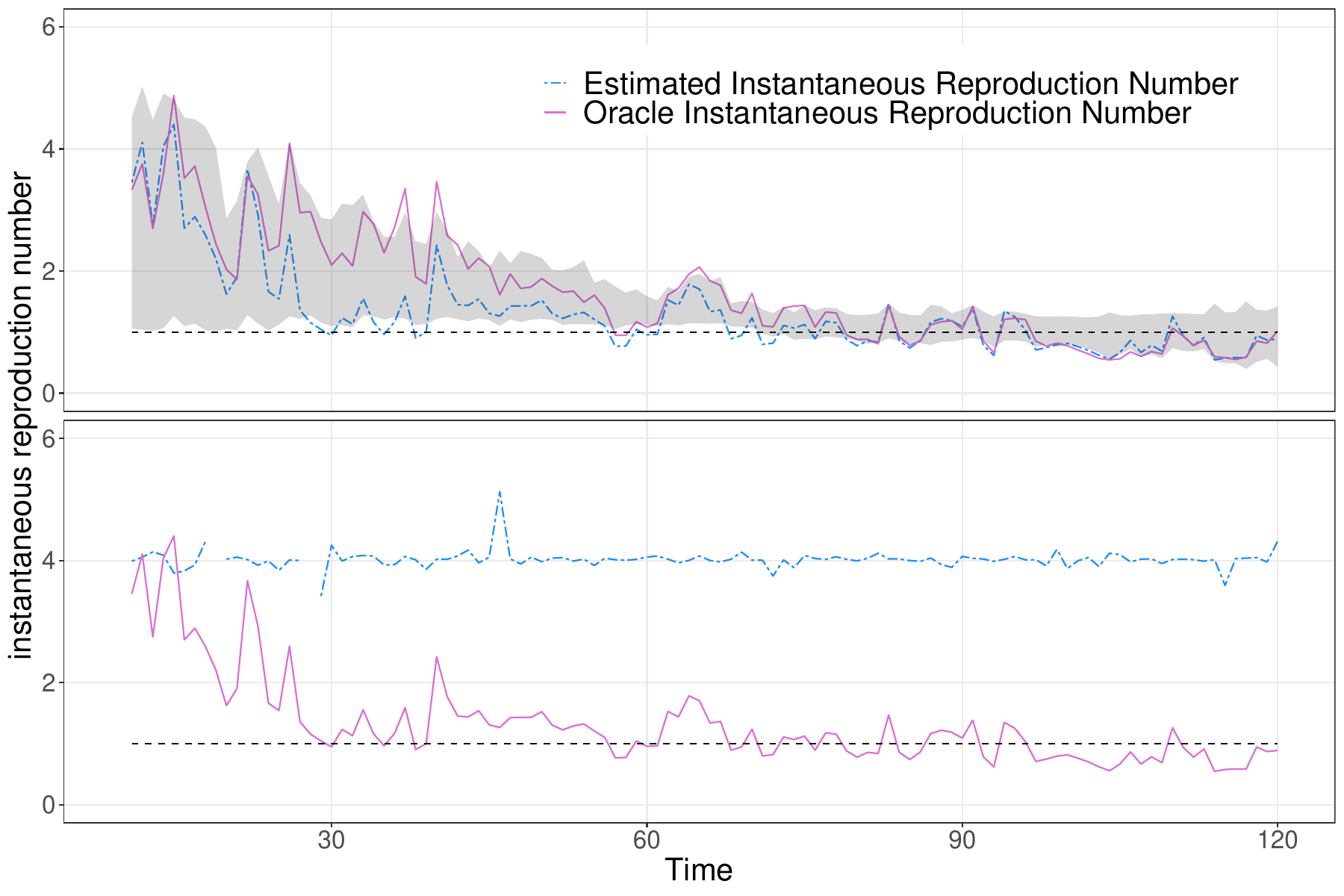}
	\caption{Robustness comparison between TSI model and basic SIR model.}
	\label{differ_model}
\end{figure}

Second, we tested the SIR model estimator proposed in \cite{chen2020time} using the data generated from the TSI model proposed in (3.2) and $I_t \sim Poisson(R_t \Lambda_t)$. Specifically,
\begin{enumerate}[Step 1.]
	\item Generate the covariates data according to the simulation section of the main manuscript.
	\item Generate the oracle time series of the instantaneous reproduction number using (3.2).
	\item Generate the incident case number from $I_t \sim Poisson(R_t \Lambda_t)$ with a initial incident case number $I_0$ and infectiousness profile according to the simulation section of the main manuscript.
	\item Generate recovery rate $\gamma_t$ from normal distribution $N(0.07,0.0025)$ and calcuate the transmission rate $\beta_t=\gamma_t \cdot R_t$.
	\item Generate the number of recovery based on the SIR equations.
	\item Estimate the transmission rate $\beta_t$ and recovery rate $\gamma_t$ using method proposed in \cite{chen2020time}.
	\item Calculate the estimation of instantaneous reproduction number by $\hat{R}_t=\hat{\beta}_t / \hat{\gamma}_t$.
\end{enumerate} 
The results are shown in the bottom panel of Figure.{A.2}. The SIR model estimator drifts away from the oracle $R_t$. Possible reasons for the performance could be that there exists measurement error in the data generation process, small sample size/incident cases number, or the sensitivity of ratio structure $\hat{R}_t=\hat{\beta}_t / \hat{\gamma}_t$ causing $\hat{R}_t$ differs from $R_t$. 

Overall, this small simulation suggests the proposed LOCAL-QUEST algorithm would have a great robust performance even under model mis-specification cases like the data are actually generated from compartment models.

\subsection*{A.5. Extension to nonparametric model}

The proposed method can be further extended to a nonparametric model by relaxing the assumption on link function to avoid misspecification. The estimation is straightforward if the dependency of $\{ R_{it}\}_{1\leq t\leq N}^{1\leq i\leq n} $ is ignored. For example, if model assumes $\e [R_{it} |Z_{it}] =f(Z_{it})$ with $\{ (R_{it},Z_{it}) \}_{t\geq 0}$ being i.i.d, then it forms a nonparametric regression problem and the function $f(\cdot)$ can be estimated using splines regression \citep{friedman1991multivariate}, wavelet regression \citep{hall1997interpolation}, or a simple Nadaraya-Watson kernel estimator. When time series dependency of $\{ R_{it}\}_{1\leq t\leq N}^{1\leq i\leq n} $ is considered, e.g., $\e \big[ R_{it} | Z_{it} , D_{3,it} \big] =\sum_{m=1}^q \theta_m f_m (R_{i,t-m}) +g(Z_t) \label{nonregression}$
with a known $R_0$ and known functions $f_m$, $1\leq m\leq q$ and unknown $g(\cdot)$, we can adopt local linear kernel estimators \citep{fan1996local} into the proposed iterative algorithm. Define
\begin{equation*}
\tilde{g}_{h,k+1}(z) = \sum_{i=1}^n \f{1}{k} \sum_{t=1}^k  W_{i,k,t,h}(z) \big(\hat{R}_{it}^{(k)}-\sum_{m=1}^q \theta_m f_m ( \hat{R}_{i,t-m}^{(k)} ) \big),
\end{equation*}
for $k\geq \tau_0$, where
\begin{align*}
W_{i,k,t,h}(z)&=\f{ \big[ s_{2,k}-s_{1,k} (Z_{it}- z) \big] K_h (Z_{it}-z) }{s_{2,k}s_{0,k}-s_{1,k}^2},\\
s_{r,k}=s_{r,k}(z)&= \sum_{i=1}^n  \f{1}{k} \sum_{t=1}^k (Z_{it}-z)^r K_h(Z_{it}-z), \;\; r=0,1,2. \\
K_h(\cdot) &= \f{1}{h} K\big( \cdot /h \big) \;\; {\rm satisfy}\; \int K_h =1. \\
\hat{R}_{it}^{(\tau_0)} & \triangleq I_{it} /\Lambda_{it}, \;\; {\rm for}\; t=1,\cdots,\tau_0, \;\; {\rm and}\;\; i=1,\cdots,n,
\end{align*}
with $h$ being the bandwidth. Consequently, for each $i=1,\cdots,n$, define 
\begin{equation*}
\tilde{R}_{it}^{(k+1)} \triangleq \sum_{m=1}^q \theta_m f_m ( \hat{R}_{i,t-m}^{(k)} ) + \tilde{g}_{h,k+1} ( Z_{it}),  \;\; {\rm for}\;\; t=\tau_0+1,\cdots,k+1.
\end{equation*}
Updates the estimator of $\theta$ based on the Quasi-score estimating equation,
\begin{equation*}
\tilde{U}_{k+1}^{*}( \theta)=\sum_{i=1}^n \sum_{t=\tau_0+1}^{k+1} \big(  \frac{ \partial \tilde{\mu}_{it}^{(k+1)} }{\partial \theta} \big)^T \big(  \tilde{\nu}_{it}^{(k+1)}    \big)^{-1} (I_{it}- \tilde{\mu}_{it}^{(k+1)})=0. \label{non-linear quasi-score equation measurement error}
\end{equation*}
with $\tilde{\mu}_{it}^{(k+1)}= \tilde{\nu}_{it}^{(k+1)}=\tilde{R}_{it}^{(k+1)} \Lambda_{it}$. Or equivalently, Updates the estimator of $\theta$ from
\begin{align}
&\hat{\theta}^{(k+1)} \triangleq \arg\max_{\Vert \theta \Vert_1<1 } \bigg[ \sum_{i=1}^n \sum_{t=\tau_0+1}^{k+1}  I_{it} \log \Big( \tilde{R}_{it}^{(k+1)} \Big) - \tilde{R}_{it}^{(t+1)} \Lambda_{it}  \bigg], \tag{A.5} \label{nonpara}
\end{align}
and obtain the estimators $\hat{R}_{it}^{(k+1)}$, $t=\tau_0+1,\cdots,k+1$ and $\hat{g}_{h,k+1}$ by plug $ \hat{\theta}^{(k+1)} $ into the corresponding $\tilde{R}_{it}^{(k+1)}$ and $\tilde{g}_{h,k+1}$. Furthermore, we have for $k\geq \tau_0$ that
%\begin{thm}[Concavity] \label{thmconcave2}
%	Equation (\ref{nonpara}) forms a globally concave maximization problem.
%\end{thm}

\begin{manualtheorem}{A.1}[Concavity] \label{thmconcave2}
	Equation (\ref{nonpara}) forms a globally concave maximization problem.
\end{manualtheorem}
Proof of this theorem can be found in Section C of the Supplementary Materials.

\section*{B. Asymptotic mixture normality of the quasi-score estimator}

	\setcounter{equation}{0}
	\renewcommand{\theequation}{B.\arabic{equation}}
	
	It's clear that on the extinction set $\mathcal{E}$, 
	\begin{equation}
	\mathcal{E}_{none}=\mathcal{E}^c,\quad {\rm and}\quad \mathcal{E} \triangleq \Big\{  \exists \;K,\; s.t., \;   I_t=0,\; {\rm for}\; t>K  \Big\} = \bigcup_{t\geq 1} \big\{  \mu_t=0 \big\}.  \label{non-extinction set} 
	\end{equation}
	no consistent estimator $\hat{\gamma}$ can be expected, and $(\hat{\gamma}-\gamma_0)^TA_n(\hat{\gamma}-\gamma_0)\rightarrow \infty$ almost surely on the extinction set if $\lambda_{\min}(A_n)\rightarrow \infty$. Unfortunately, due to the high variability introduced by different $R_t$ and $x_t$, there is no neat form for calculating the extinction probability unlike the case in Bienaym\'{e}-Galton-Watson branching process. So we simply left the extinction probability
	\begin{equation*}
	\P\big(  \mathcal{E}   \big) = \P\bigg(  \bigcup_{t\geq q} \{  \mu_t=0 \}      \bigg)
	\end{equation*}
	there and only focusing on the asymptotic behaviour of $\hat{\gamma}$ on the non-extinction set $\mathcal{E}_{none}$. 
	
	%To study the asymptotic property of $\hat{\gamma}$, we focus our investigation on the non-extinction set $\mathcal{E}_{none}$ to avoid finite sample behavior which is difficult to track from case to case,
	
	According to theorem 3 of \cite{kaufmann1987regression},  and assume the following condition \ref{condition quasi},
	\begin{condition}\label{condition quasi}
		\begin{enumerate}[(i).]
			\item There exists some nonrandom nonsingular normalizing matrix $A_N$, s.t., the normalized conditional variance converge to a a.s. positive definite random matrix $\zeta^T \zeta$, i.e.,
			\begin{equation*}	 
			A_N^{-1} \Big[ \sum_{t=q}^{N}\cov  \big(   \xi_t (\gamma_0) | \mathcal{F}_{t-1}    \big)	  \Big] \big( A_N^{-1} \big)^T \overset{P}{\longrightarrow} \zeta^T \zeta.
			\end{equation*}
			\item The conditional Lindeberg condition holds, i.e., for all $\epsilon>0$, 
			\begin{equation*}
			\sum_{t=q}^N \e \Big[ \xi_t^T(\gamma_0) \big(A_N^T A_N\big)^{-1} \xi_t(\gamma_0) \cdot \mathbbm{1}(|\xi_t^T(\gamma_0)  \big(A_N^T A_N\big)^{-1} \xi_t(\gamma_0)|>\epsilon^2) \big| \mathcal{F}_{t-1}  \Big] \overset{P}{\longrightarrow} 0,
			\end{equation*}
			\item The smoothness condition
			\begin{equation*}
			\sup_{\tilde{\gamma}\in \mathcal{B}_N (\delta)} \bigg \Vert  A_N^{-1} \Big(  \f{\partial U_N(\tilde{\gamma})}{\partial \gamma}	+ \sum_{t=q}^{N}\cov  \big(   \xi_t(\gamma_0) | \mathcal{F}_{t-1}    \big)	  \Big)   \big( A_N^{-1} \big)^T     \bigg \Vert \overset{P}{\longrightarrow} 0,
			\end{equation*}
			with $\mathcal{B}_N (\delta)= \{ \tilde{\gamma}:  \Vert A_N^T (\tilde{\gamma} - \gamma_0) \Vert \leq \delta  \}$, holds for all $\delta>0$.
		\end{enumerate}
	\end{condition}
	we have the asymptotic normality of $\hat{\gamma}$ under certain normalization. While the (iii) of condition \ref{condition quasi} is hard to track and unnecessary for the special case: the embedded autocorrelated latent process ({3.2}), as an alternative, we propose the following condition \ref{Condition for mixture normal}. Define
	\begin{equation*}
	y_t=\log (R_t) - Z_t^T \beta = \log (\mu_t)-\log(\Lambda_t) - Z_t^T \beta,
	\end{equation*}
	then $\{ y_t \}_{t\geq 0}$ forms a degenerated AR$(q)$ model or a recurrence equation with order $q$. To avoid divergence, we require $M=\max_{t} \Vert Z_t \Vert_2<\infty$ and $\{ y_t \}_{t\geq 0}$ to be causal, i.e, the roots of its characteristic polynomial are outside the unit circle. 
	
	\begin{condition}\label{Condition for mixture normal}
		\begin{enumerate}[(i).]
			\item Assume (i) and (ii) of condition \ref{condition quasi} holds for $A_N^2=\sum_{t=q}^{N} Cov(\xi_t(\gamma_0))$. \label{CM1}
			\item Assume $\{ A_t^{-1} V_t^2(\gamma_0) (A_t^{-1} )^T \}_{t\geq 1}$ defined in (\ref{Jacobian terms}) is termwise uniformly integrable. \label{CM2}
			\item On $\mathcal{E}_{none}$, for $\forall \gamma \in \Theta$, the minimum eigenvalue of the normalized matrix
			\begin{equation*}
			I(\xi) \triangleq A_N^{-1}  \left( - \f{\partial U_N(\gamma)}{\partial \gamma}\Big|_{\gamma=\xi }   \right) (A_N^{-1})^T 
			\end{equation*}
			is uniformly bounded away from $0$, i.e, $\exists \lambda_0>0$, s.t., $\lambda_{\min}(I(\xi)) \geq \lambda_0$. \label{CM3}
			\item There exist some $\phi$, s.t, $\Vert \theta(\gamma_0) \Vert_1 < \phi <1$,
			$
			\lim  \big(   \lambda_{\min}(A_N)  \big)^{-1} \sum_{t=q}^{N}  \phi^{[t/q]} \e \mu_t <+\infty.
			$ \label{CM4}
		\end{enumerate}
	\end{condition}
	%\begin{thm}
	\begin{manualtheorem}{B.1}
		\label{thm mixture normal}
		Under condition \ref{Condition for mixture normal}, $ 
		\Big[  \sum_{t=q}^{N}\cov  \big(   \xi_t (\gamma_0) | \mathcal{F}_{t-1}    \big)  \Big]^{1/2} (\hat{\gamma}-\gamma_0) \overset{d}{\longrightarrow} \mathcal{N}(0,I).$
	%\end{thm}
	\end{manualtheorem}
	
	\begin{proof}
		(\emph{of Theorem \ref{thm mixture normal}})
		Without loss of generality, assume $I_0=1$ and $\{ R_t, 0\leq t< q \}$ are known. Since $\theta_m>0$ and $\{ y_t \}_{t\geq 0}$ is causal, so $\Vert \theta \Vert_1 = \sum_{m=1}^q \vert \theta_m  \vert <1$.	By (3.2), for $t\geq q$,
		\begin{align}
		\frac{\partial \log(\mu_t)  }{\partial \beta} &= Z_t^T + \frac{\partial  y_t }{\partial \beta}, \quad 
		\frac{\partial y_t}{\partial \beta} = \sum_{m=1}^{q} \theta_m \frac{\partial y_{t-m}}{\partial \beta}, \quad \f{\partial y_{s}}{\partial \beta}=-Z_s^T, \;\; {\rm for}\;\; 0\leq s<q,  \nn  \\
		\frac{\partial \log(\mu_t)  }{\partial \theta} &=\frac{\partial  y_t }{\partial \theta}= (y_{t-1},\cdots , y_{t-q})+\sum_{m=1}^{q} \theta_m \f{ \partial y_{t-m}  }{\partial \theta} , \quad 
		\f{\partial y_{s}}{\partial \theta} =0, \;\; {\rm for}\;\; 0\leq s<q.  \label{recurrence equation one}
		\end{align}
		Notice that $R_t$, as well as $y_t$, is determined when given $\beta$ and $\theta$, hence $ \partial \log\big(  \mu_t  \big) / \partial \gamma$ is non-random when given $\beta$ and $\theta$.
		
		Recall that $\hat{\gamma}$ is the solution of the equation $U_N(\gamma)=0$ and $\gamma_0$ is the oracle parameter value. Thereby, for some $\xi$ depends on $\{ I_t, t\geq 0 \}$ and lays between $\gamma_0$ and $\hat{\gamma}$, we have
		\begin{align*}
		0=U_N(\hat{\gamma})=U_N(\gamma_0)+\f{\partial U_N(\gamma)}{\partial \gamma}\Big|_{\gamma=\xi } (\hat{\gamma}-\gamma_0),
		\end{align*}
		or equivalently, 
		\begin{align*}
		A_N^{-1}  \left( T_{N,1}(\xi)+T_{N,2}(\xi) + V_N^2(\xi)   \right) (A_N^{-1 })^T  A_N^T (\hat{\gamma}-\gamma_0) &= A_N^{-1} U_N(\gamma_0) ,
		\end{align*}
		where $A_N^T A_N=\sum_{t=q}^N Cov(\xi_t(\gamma_0))= \e V_N^2(\gamma_0)$, so without loss of generality, we may choose $A_N$ to be symmetric,
		\begin{align}
		&T_{N,1}(\xi)= - \sum_{t=q}^{N}  \frac{\partial \log(\mu_t)  }{\partial \gamma \partial \gamma^T} \Big|_{\gamma=\xi} (I_t-\mu_t |_{\gamma_0}), \nn \\
		&T_{N,2}(\xi)= \sum_{t=q}^{N}  \frac{\partial \log(\mu_t)  }{\partial \gamma \partial \gamma^T} \Big|_{\gamma=\xi} (\mu_t|_{\xi}-\mu_t|_{\gamma_0}), \nn \\
		\quad {\rm and}\quad &V_N^2(\xi)= \sum_{t=q}^{N}   \left( \f{\partial \log(\mu_t) }{\partial \gamma } \right)^T   \frac{\partial \log(\mu_t)  }{\partial \gamma }\mu_t\Big|_{\gamma=\xi}  . \label{Jacobian terms}
		\end{align}
		To clear the confusion, when taking expectation or mentioning terms without specifically marked in the later paragraph, the $I_t$ and $\mu_t$ would always refer to the random variables generated with oracle parameters $\gamma_0$.
		
		For an arbitrary given and fixed vector $\alpha=(\alpha_1,\cdots,\alpha_{p+q})^T \in \mathcal{M}_{(p+q)\times 1}$, define
		\begin{align*}
		S_N(\alpha) \triangleq  \f{ \alpha^T}{\Vert \alpha \Vert} A_N^{-1} U_N(\gamma_0) &= \sum_{t=q}^{N} \f{ \alpha^T}{\Vert \alpha \Vert} A_N^{-1} \left( \frac{\partial \log(\mu_t)  }{\partial \gamma} \right)^T (I_t- \mu_t|_{\gamma_0})  \triangleq \sum_{t=q}^{N} \tilde{\xi}_t .
		\end{align*}
		It's trivial to see that $\tilde{\xi}_t\in \mathcal{F}_t$ and $\e (\tilde{\xi}_t | \mathcal{F}_{t-1})=0$, hence $\{  \tilde{\xi}_t,t\geq q\}$ is a martingale difference sequence, and as one may expect, the behavior of the conditional variance
		\begin{align*}
		\tilde{V}_N^2 (\alpha) &\triangleq  \sum_{t=q}^{N} \e \big(  \tilde{\xi}_t^2 | \mathcal{F}_{t-1}    \big)=\f{ \alpha^T}{\Vert \alpha \Vert} A_N^{-1} V_N^2(\gamma_0) ( A_N^{-1} )^T  \f{ \alpha}{\Vert \alpha \Vert}
		\end{align*}
		plays an important role and would directly affect the limiting distribution of $S_N(\alpha)$. Condition \ref{Condition for mixture normal} (i) automatically ensures that 
		there exist some random variable $\zeta(a)= ( \f{\alpha^T}{\Vert \alpha \Vert } \zeta^T \zeta \f{\alpha}{\Vert \alpha \Vert } )^{1/2}$, s.t., $\e \zeta^2(\alpha) <\infty$ and 
		\begin{equation}
		\tilde{V}_N^2(\alpha) \overset{P}{\longrightarrow} \zeta^2(\alpha). \label{condition variance convergence}
		\end{equation}
		Meanwhile, (\ref{CM2}) of condition \ref{Condition for mixture normal} implies, for $\forall \alpha$,
		\begin{equation*}
		\Big\{\tilde{V}_t^2(\alpha) \Big\}_{t\geq 1} \in Convex \Big\{ \Big( A_t^{-1} V_t^2(\gamma_0) ( A_t^{-1} )^T \Big)_{js}, 1\leq j,s \leq p+q, t\geq 1  \Big\}
		\end{equation*}
		is uniformly integrable, so we conclude $\tilde{V}_N^2(\alpha) \overset{L_1}{\longrightarrow} \zeta^2(\alpha)$ and $\e  \tilde{V}_N^2 (\alpha) = \e  \zeta^2(\alpha)  =1 $. Hence $\e (\max_{t\geq q} \tilde{\xi}_t^2)\leq1$ uniformly in $N$. Together with the conditional Lindeberg condition (\ref{CM1}) in condition \ref{Condition for mixture normal}, we conclude $\tilde{U}_N^2 (\alpha) - \tilde{V}_N^2(\alpha) \overset{L_1}{\longrightarrow} 0$ by theorem 2.23 of \cite{hall2014martingale} for $\tilde{U}_N^2 (\alpha)\triangleq \sum_{t=q }^N \tilde{\xi}_t^2$, hence $\tilde{U}_N^2(\alpha) \overset{L_1}{\longrightarrow} \zeta^2(\alpha)$. 
		
		Further, since
		\begin{equation*}
		0\leq  \sum_{t=q}^N \e \Big[ \tilde{\xi}_t^2 \mathbbm{1}(|\tilde{\xi}_t|>\epsilon) \big| \mathcal{F}_{t-1}  \Big] \leq \tilde{V}_N^2 (\alpha)= \sum_{t=q}^N \e \big[ \tilde{\xi}_t^2 \big| \mathcal{F}_{t-1}  \big], \;\; a.s., \;\; {\rm for\; all\; }N\geq 1,
		\end{equation*}
		so by a variation of dominant convergence theorem  (theorem 1 in \cite{pratt1960interchanging}), we have the Conditional Lindeberg condition is equivalent to the Lindeberg condition here, i.e., 
		\begin{gather}
		{\rm for \; all \;} \epsilon>0, \quad \sum_{t=q}^N \e \Big[ \tilde{\xi}_t^2 \mathbbm{1}(|\tilde{\xi}_t|>\epsilon) \Big] \rightarrow 0, \quad {\rm which \; further \; implies}  \nn \\
		\Rightarrow\;\; {\rm for \; all \;} \epsilon>0, \quad   \sum_{t=q}^N  \tilde{\xi}_t^2 \mathbbm{1}(|\tilde{\xi}_t|>\epsilon)  \overset{P}{\rightarrow} 0,  \nn \\
		\Leftrightarrow\;\; {\rm for \; all \;} \epsilon>0, \quad \P\Big(     \sum_{t=q}^N \tilde{\xi}_t^2 \mathbbm{1}(|\tilde{\xi}_t|>\epsilon)  > \epsilon^2     \Big)=\P\Big(\max_{t\geq q} |\tilde{\xi}_t|>\epsilon \Big)  \rightarrow 0,  \nn \\
		\Leftrightarrow\;\; {\rm for \; all \;} \epsilon>0, \quad \max_{t\geq q} |\tilde{\xi}_t|  \overset{P}{\rightarrow} 0. \label{negligable}
		\end{gather}
		Combine (\ref{negligable}), the fact that $\tilde{U}_N^2(\alpha) \overset{L_1}{\longrightarrow} \zeta^2(\alpha)$ and $\e (\max_{t\geq q} \tilde{\xi}_t^2)\leq1$ uniformly in $N$, we claim that by using the martingale central limit theorem (see e.g, theorem 3.2 of \cite{hall2014martingale}), for $Z$ being a standard normal distributed random variable and being independent of $\zeta(\alpha)$,
		\begin{equation*}
		S_N(\alpha) \overset{d}{ \rightarrow} \zeta(\alpha) \cdot Z \;\; ({\rm stably}).
		\end{equation*}
		Hence the $A_N^{-1}U_N(\gamma_0)$ converge to certain distribution with characteristic function
		\begin{equation*}
		f_{A_N^{-1}U_N(\gamma)} (\alpha)  =\e \exp\Big(  i\alpha ^T A_N^{-1} U_N(\gamma_0)   \Big) \rightarrow \e \exp \Big(  -\f{1}{2} \Vert \alpha \Vert_2^2 \zeta^2(\alpha)      \Big),
		\end{equation*}
		or equivalently,
		\begin{equation}
		A_N^{-1}U_N(\gamma_0) \overset{d}{ \rightarrow} \zeta^T \cdot Z  \;\; ({\rm stably}).  \label{distribution}
		\end{equation}
		for $\zeta$ independent of $Z\sim \mathcal{N}(0,I)$. Of course, the limit distribution holds only on the non-extinction set $\mathcal{E}_{none}$. Since $\zeta$, $Z$ are also measurable, so we have $A_N^{-1}U_N(\gamma_0) \overset{P}{ \rightarrow} \zeta^T \cdot Z$.
		
		Next, for the $(p+q)\times (p+q)$ matrix $\partial U_N(\gamma)/\partial \gamma$, 
		apparently, we have $V_N^2(\xi)$ being non-negative definite while $T_{N,1}(\gamma_0)$ is a martingale sequence with mean zero. According to (\ref{CM3}) of condition \ref{Condition for mixture normal}, for $\forall \alpha\in \mathcal{M}_{(p+q)\times 1}$ and $\alpha \neq 0$,
		\begin{equation}
		\bigg|  \f{\alpha^T }{\Vert \alpha \Vert} A_N (\hat{\gamma}-\gamma_0)  \bigg| = \bigg| \f{\alpha^T }{\Vert \alpha \Vert} A_N^T  \left( - \f{\partial U_N(\gamma)}{\partial \gamma}\Big|_{\gamma=\xi }   \right)^{-1} A_N A_N^{-1} U_N(\gamma_0) \bigg| <\infty, \;\; a.s.  \label{consistency}
		\end{equation} 
		Meanwhile, from (\ref{negligable}), we know that conditional Lindeberg condition implies that for all $t\geq q$ and all $\alpha$,
		\begin{align}
		\f{ \alpha^T}{\Vert \alpha \Vert} A_N^{-1}\left( \frac{\partial \log(\mu_t)  }{\partial \gamma} \right)^T  \left( \frac{\partial \log(\mu_t)  }{\partial \gamma} \right) \bigg|_{\gamma=\gamma_0} (A_N^{-1})^T \f{ \alpha}{\Vert \alpha \Vert}  \mu_t \; \overset{P}{\longrightarrow} \; 0. \label{sn}
		\end{align}
		$V_N^2(\gamma_0)$ being positive definite for large $N$ implies that on the non-extinction set, there exists $\{ k_1,\cdots,k_q \}$ such that $ \{ \partial \log (\mu_{t})/\partial \gamma, t\in \{ k_1,\cdots,k_q \}  \}$ are linear independent, combine with (\ref{sn}) concludes that the minimum eigenvalue of $A_N$ increase to infinity, denote as $\lambda_{\min}(A_N) \rightarrow +\infty$. Thus (\ref{consistency}) implies $\hat{\gamma}-\gamma_0 \overset{P}{\rightarrow} 0$.
		
		Based on (\ref{recurrence equation one}), we have
		\begin{gather}
		\frac{\partial \log(\mu_t)  }{\partial \beta \partial \beta^T}=0, \quad   \frac{\partial \log(\mu_t)  }{\partial \beta \partial \theta^T}=\Big( \big( \frac{\partial y_{t-1}  }{\partial \beta} \big)^T, \cdots, \big( \frac{\partial y_{t-q}  }{\partial \beta} \big)^T \Big)+ \sum_{m=1}^q \theta_m \f{\partial y_{t-m}}{\partial \beta \partial \theta^T}, \nn \\
		\frac{\partial \log(\mu_t)  }{\partial \theta \partial \theta^T }=  \frac{\partial y_t  }{\partial \theta \partial \theta^T } =2\Big( \big( \frac{\partial y_{t-1}  }{\partial \theta} \big)^T, \cdots, \big( \frac{\partial y_{t-q}  }{\partial \theta} \big)^T \Big)^T + \sum_{m=1}^q \theta_m \f{\partial y_{t-m}}{\partial \theta \partial \theta^T} .  \label{recurrence equation two}
		\end{gather}
		By denote the bound of the first $q$ terms of $y_t|_{\gamma_0}$ and $\partial y_t /\partial \beta_j |_{\gamma_0}$, $1\leq j\leq p$, as $\tilde{M}$, then it's straightforward to see from induction that for all $t$,
		\begin{equation*}
		\bigg|  \f{\partial y_{t}}{ \partial \beta_j} \big|_{\gamma_0} \bigg| \leq \Vert \theta (\gamma_0) \Vert_1^{[t/q]} \tilde{M}, \;\; 1\leq j\leq p, \quad {\rm and}\quad  \Big|  y_t\big|_{\gamma_0} \Big|   \leq \Vert \theta (\gamma_0) \Vert_1^{[t/q]} \tilde{M}
		\end{equation*}
		Here, $[\;\cdot\;]$ denotes the floor function. Since $\Vert \theta (\gamma_0) \Vert_1<1$, define 
		\begin{equation*}
		\tilde{M}' = \f{\tilde{M}/ \Vert \theta (\gamma_0) \Vert_1 }{\phi- \Vert \theta (\gamma_0) \Vert_1 }+\max_{1\leq m\leq q, 0\leq t<q} \bigg\{    \bigg|  \f{\partial y_{t}}{ \partial \theta_m} \big|_{\gamma_0} \bigg|   \bigg\},
		\end{equation*}
		so for $1\leq m\leq q$, we have
		\begin{align*}
		\bigg|  \f{\partial y_{t}}{ \partial \theta_m} \big|_{\gamma_0} \bigg| &= \bigg| y_{t-m}\big|_{\gamma_0}  + \sum_{s=1}^q \theta_s \f{\partial y_{t-s}}{\partial \theta_m}\big|_{\gamma_0}  \bigg| \leq \Big| y_{t-m}\big|_{\gamma_0}   \Big| + \Vert \theta (\gamma_0) \Vert_1 \max_{1\leq s\leq q} \bigg\{    \bigg|  \f{\partial y_{t-s}}{ \partial \theta_m} \big|_{\gamma_0} \bigg|   \bigg\} \\
		& \leq \Vert \theta (\gamma_0) \Vert_1^{1+[(t-m)/q]} \f{\tilde{M}}{  \Vert \theta (\gamma_0) \Vert_1  } + \Vert \theta (\gamma_0) \Vert_1 \max_{1\leq s\leq q} \bigg\{    \bigg|  \f{\partial y_{t-s}}{ \partial \theta_m} \big|_{\gamma_0} \bigg|   \bigg\} \\
		& \leq  \Big( \phi- \Vert \theta (\gamma_0) \Vert_1 \Big) \phi^{[t/q]}  \tilde{M}'  + \Vert \theta (\gamma_0) \Vert_1 \max_{1\leq s\leq q} \bigg\{    \bigg|  \f{\partial y_{t-s}}{ \partial \theta_m} \big|_{\gamma_0} \bigg|   \bigg\}
		\end{align*}
		then it's straightforward to see from induction that for all $t$,
		\begin{equation*}
		\bigg|  \f{\partial y_{t}}{ \partial \theta_m} \big|_{\gamma_0} \bigg| \leq \phi^{[t/q]} \tilde{M}', \;\; 1\leq m\leq q.
		\end{equation*}
		Now, for $0\leq t<q$, it's obvious from (3.2) and (\ref{recurrence equation one}) that
		\begin{align}
		\bigg \vert \frac{\partial y_t  }{\partial \beta_j} \big|_{\xi} -  \frac{\partial y_t  }{\partial \beta_j} \big|_{\gamma_0} \bigg \vert  &= 0 \leq \phi^{[t/q]} M \cdot\sqrt{q}\Vert \xi-\gamma_0 \Vert_2, \quad {\rm for}\;\; 1\leq j\leq p, \nn \\
		\bigg \vert \frac{\partial y_t  }{\partial \theta_m} \big|_{\xi} -  \frac{\partial y_t  }{\partial \theta_m} \big|_{\gamma_0} \bigg \vert  &= 0 \leq \phi^{[t/q]} M'\cdot \sqrt{q}\Vert \xi-\gamma_0 \Vert_2, \quad {\rm for}\;\; 1\leq m\leq q, \nn \\
		\Big \vert y_t  \big|_{\xi} -  y_t  \big|_{\gamma_0} \Big \vert   &\leq  \phi^{[t/q]} M \cdot \sqrt{q} \Vert \xi-\gamma_0 \Vert_2,      \label{induction}
		\end{align}
		for finite 
		\begin{equation*}
		M=\max \Big\{ \f{3\tilde{M}}{\phi-\Vert \theta(\gamma_0) \Vert_1}, \Vert Z_t \Vert_2, t\geq 0 \Big\}, \quad {\rm and}\quad M'= \f{3(\tilde{M}'+M)}{\phi-\Vert \theta(\gamma_0) \Vert_1}.
		\end{equation*}
		Suppose (\ref{induction}) holds for all $0\leq t<k$, then for $t=k$ and $\sqrt{q}\Vert \xi -\gamma_0 \Vert_2< (\phi-\Vert \theta(\gamma_0) \Vert_1)/3$, which will be the case for large $N$ and with high probability since $\xi$ lays between $\gamma_0$ and its consistent estimator $\hat{\gamma}$,
		\begin{align*}
		&\bigg \vert \frac{\partial y_k  }{\partial \beta_j} \big|_{\xi} -  \frac{\partial y_k  }{\partial \beta_j} \big|_{\gamma_0} \bigg \vert  =  \bigg \vert  \sum_{m=1}^q \big( \theta_m \frac{\partial y_{k-m}  }{\partial \beta_j} \big) \big|_{\xi} -   \sum_{m=1}^q \big( \theta_m \frac{\partial y_{k-m}  }{\partial \beta_j} \big) \big|_{\gamma_0}  \bigg \vert   \\
		\leq &\bigg \vert  \sum_{m=1}^q \big( \theta_{m,\xi}-\theta_{m,\gamma_0}   \big) \big( \frac{\partial y_{k-m}  }{\partial \beta_j}  \big|_{\xi}   -  \frac{\partial y_{k-m}  }{\partial \beta_j}  \big|_{\gamma_0}  \big) \bigg \vert    +  \bigg \vert     \sum_{m=1}^q \theta_{m,\gamma_0} \big( \frac{\partial y_{k-m}  }{\partial \beta_j}\big|_{\xi}- \frac{\partial y_{k-m}  }{\partial \beta_j}\big|_{\gamma_0}  \big)  \bigg \vert   \\
		& \qquad +  \bigg \vert  \sum_{m=1}^q \big( \theta_{m,\xi}-\theta_{m,\gamma_0}   \big) \frac{\partial y_{k-m}  }{\partial \beta_j}  \big|_{\gamma_0}   \bigg \vert   \\
		\leq & \bigg \vert  \sum_{m=1}^q \big( \theta_{m,\xi}-\theta_{m,\gamma_0}   \big)  \bigg \vert \cdot \phi^{[k/q]-1} M\sqrt{q} \Vert \xi-\gamma_0 \Vert_2   +  \bigg \vert     \sum_{m=1}^q \theta_{m,\gamma_0}  \bigg \vert  \cdot \phi^{[k/q]-1}M \sqrt{q} \Vert \xi-\gamma_0 \Vert_2  \\
		& \qquad +  \bigg \vert  \sum_{m=1}^q \big( \theta_{m,\xi}-\theta_{m,\gamma_0}   \big)  \bigg \vert  \cdot  \Vert \theta (\gamma_0) \Vert_1^{[k/q]-1} \tilde{M}  \leq  \phi^{[k/q]} M \cdot \sqrt{q} \Vert \xi-\gamma_0 \Vert_2
		\end{align*}
		holds for $1\leq j\leq p$, and similarly, we have
		\begin{equation*}
		\Big \vert y_k  \big|_{\xi} -  y_k  \big|_{\gamma_0} \Big \vert    \leq  \phi^{[k/q]} M \cdot \sqrt{q} \Vert \xi-\gamma_0 \Vert_2. 
		\end{equation*}
		Correspondingly, for $1\leq m\leq q$,
		\begin{align*}
		&\bigg \vert \frac{\partial y_k  }{\partial \theta_m} \big|_{\xi} -  \frac{\partial y_k  }{\partial \theta_m} \big|_{\gamma_0} \bigg \vert  \leq   \bigg \vert  y_{k-m} \big|_{\xi} - y_{k-m} \big|_{\gamma_0}    \bigg\vert + \bigg\vert \sum_{s=1}^q \big( \theta_s \frac{\partial y_{k-s}  }{\partial \theta_m} \big) \big|_{\xi} -   \sum_{s=1}^q \big( \theta_s \frac{\partial y_{k-s}  }{\partial \theta_m} \big) \big|_{\gamma_0}  \bigg \vert   \\
		\leq & \bigg \vert  \sum_{s=1}^q \big( \theta_{s,\xi}-\theta_{s,\gamma_0}   \big)  \bigg \vert \cdot \phi^{[k/q]-1} M'\sqrt{q} \Vert \xi-\gamma_0 \Vert_2   +  \bigg \vert     \sum_{s=1}^q \theta_{s,\gamma_0}  \bigg \vert  \cdot \phi^{[k/q]-1} M' \sqrt{q} \Vert \xi-\gamma_0 \Vert_2  \\
		& \;\; +  \bigg \vert  \sum_{s=1}^q \big( \theta_{s,\xi}-\theta_{s,\gamma_0}   \big)  \bigg \vert  \cdot  \phi^{[k/q]-1} \tilde{M}'  + \phi^{[(k-i)/q]} M \sqrt{q} \Vert \xi-\gamma_0 \Vert_2  \leq \phi^{[k/q]}M'\cdot \sqrt{q} \Vert \xi-\gamma_0 \Vert_2
		\end{align*}
		Hence by induction that (\ref{induction}) holds for all $t\geq 0$. Similarly, we may use induction to conclude that 
		\begin{gather}
		\Big\vert \frac{\partial \log(\mu_t)  }{\partial \beta_j \partial \theta_{m_1} } \big|_{\gamma_0} \Big \vert  \leq \phi^{[t/q]} \tilde{M}'', \;\;
		\Big\vert \frac{\partial \log(\mu_t)  }{\partial \beta_j \partial \theta_{m_1} } \big|_{\xi} - \frac{\partial \log(\mu_t)  }{\partial \beta_j \partial \theta_{m_1} } \big|_{\gamma_0}  \Big \vert  \leq \phi^{[k/q]} M'' \cdot \sqrt{q} \Vert \xi-\gamma_0 \Vert_2,  \nn \\
		\Big\vert \frac{\partial \log(\mu_t)  }{\partial \theta_{m_1} \partial \theta_{m_2} } \big|_{\gamma_0} \Big \vert  \leq   \phi^{[t/q]} \tilde{M}'', \;\;    \Big\vert \frac{\partial \log(\mu_t)  }{\partial \theta_{m_1} \partial \theta_{m_2} } \big|_{\xi} - \frac{\partial \log(\mu_t)  }{\partial \theta_{m_1} \partial \theta_{m_2} } \big|_{\gamma_0}  \Big \vert  \leq \phi^{[k/q]} M'' \cdot \sqrt{q} \Vert \xi-\gamma_0 \Vert_2, \label{induction two}
		\end{gather}
		for $1\leq j \leq p$, $1\leq m_1,m_2\leq q$ and some constants $M'', \tilde{M}''$ depends on $\Vert \theta(\gamma_0 ) \Vert_1 $ but not $t$. 
		%	Notice here the $\phi$ can be replaced by any number greater than $\Vert \theta(\gamma_0) \Vert_1$ and less than 1, the above bounds remain hold.
		
		Since $A_N^{T}A_N$ is symmetric and positive definite, so there exist a orthonormal matrix $P$ such that $A_N^{T}A_N=P^T \Lambda P$, where $\Lambda=$diag$(\lambda_1,\cdots,\lambda_{p+q})$ with $|\lambda_{p+q}| \geq \cdots \geq |\lambda_{1}|$. Accordingly, we may choose a symmetric $A_N$ such that $A_N^{-1} = P^T \Lambda^{-1/2} P$, by denote the Frobenius norm as $\Vert \cdot \Vert_{F}$, we have
		\begin{align}
		&\big\Vert A_N^{-1} \Big( T_{N,1} (\xi) - T_{N,1} (\gamma_0) \Big)  (A_N^{-1})^T \big\Vert_F \nn  \\
		\leq & \sum_{t=q}^{N} \Vert \Lambda^{-1/2} \Vert_F^2 \cdot   \Vert P \Vert_F^4  \cdot \bigg\Vert \frac{\partial \log(\mu_t)  }{\partial \gamma \partial \gamma^T } \big|_{\gamma_0} - \frac{\partial \log(\mu_t)  }{\partial \gamma \partial \gamma^T } \big|_{\xi} \bigg \Vert_F \cdot  |I_t-\mu_t|\big|_{\gamma_0} \nn  \\
		\leq & \sum_{t=q}^{N} \Big( \sum_{k=1}^{p+q} \f{1}{\lambda_k}   \Big) \cdot (p+q)^2 \cdot \sqrt{2pq+q^2} \max_{ \substack{1\leq k\leq (p+q),\\ 1\leq m\leq q} }  \Big\vert \frac{\partial \log(\mu_t)  }{\partial \gamma_k \partial \theta_m } \big|_{\xi} - \frac{\partial \log(\mu_t)  }{\partial \gamma_k \partial \theta_m } \big|_{\gamma_0}  \Big \vert  \cdot  |I_t-\mu_t| \big|_{\gamma_0}  \nn  \\
		\leq & (p+q)^3 \Big( \sum_{k=1}^{p+q} \f{1}{\lambda_k}   \Big) \sum_{t=q}^{N}  \phi^{[t/q]} M'' \cdot \sqrt{q} \Vert \hat{\gamma}-\gamma_0 \Vert_2 \big( I_t+\mu_t\big|_{\gamma_0} \big). \label{Tn difference}
		\end{align}
		So under the (\ref{CM4}) of condition \ref{Condition for mixture normal}, we have, for $\forall \epsilon>0$, 
		\begin{align}
		&\P\Big(   \big\Vert A_N^{-1} \Big( T_{N,1} (\xi) - T_{N,1} (\gamma_0) \Big)  (A_N^{-1})^T \big\Vert_F >\epsilon    \Big) \nn \\
		\leq & \P  \left(   \Vert \hat{\gamma}-\gamma_0 \Vert_2 \cdot   \big(   \lambda_{\min}(A_N)  \big)^{-1} \sum_{t=q}^{N}  \phi^{[t/q]} (I_t +\mu_t) >\f{\epsilon}{\sqrt{q} (p+q)^4 M''}  \right) \nn \\
		\leq & \P \Big(  \Vert \hat{\gamma}-\gamma_0 \Vert_2 >\delta \Big)+ \f{2\delta \sqrt{q} (p+q)^4 M''}{\epsilon} \big(   \lambda_{\min}(A_N)  \big)^{-1} \sum_{t=q}^{N}  \phi^{[t/q]} \e \mu_t  \rightarrow 0 \label{Tn difference prob}
		\end{align}
		by let $\delta$ goes to $0$ and $N$ goes to $\infty$. Hence 
		\begin{equation}
		A_N^{-1}  \big( T_{N,1} (\xi) - T_{N,1} (\gamma_0) \big)  (A_N^{-1})^T\overset{P}{\longrightarrow} 0_{(p+q)\times (p+q)}. \label{1}
		\end{equation}
		Meanwhile, under the event $\{ \Vert \hat{\gamma} - \gamma_0 \Vert_2 \leq \delta\}$ for some $\delta\leq 1$, 
		\begin{equation*}
		\bigg\vert \log \Big[ \f{R_t(\xi)}{R_t(\gamma_0)} \Big]  \bigg\vert \leq   \Big \vert y_t  \big|_{\xi} -  y_t  \big|_{\gamma_0} \Big \vert + \Vert Z_t \Vert_2 \cdot \big\Vert \beta |_{\xi} - \beta|_{\gamma_0}  \big\Vert \leq (1+\sqrt{q}) \delta M 
		\end{equation*}
		hence 
		\begin{equation}
		\bigg\vert\f{R_t(\xi)}{R_t(\gamma_0)}-1  \bigg\vert \leq \exp\Big[ (1+\sqrt{q})  M \Big] (1+\sqrt{q})M \delta \triangleq \bar{M} \delta . \label{difference between mu_t}
		\end{equation}
		On one hand, if we define an intermediate term $V_{N,int}^2$ as 
		\begin{equation*}
		V_{N,int}^2= \sum_{t=q}^{N}   \left( \f{\partial \log(\mu_t) }{\partial \gamma } \right)^T   \frac{\partial \log(\mu_t)  }{\partial \gamma }\Big|_{\gamma_0} \cdot \mu_t\big|_{\gamma=\xi},
		\end{equation*} 
		then
		\begin{equation*}
		\big\Vert A_N^{-1} \Big( V_{N,int}^2 - V_N^2 (\gamma_0) \Big)  (A_N^{-1})^T \big\Vert_F  \leq \bar{M} \delta \cdot \big\Vert A_N^{-1} V_N^2 (\gamma_0)  (A_N^{-1})^T\big\Vert_F,
		\end{equation*}
		which leads to 
		\begin{align}
		&\P\Big(   \big\Vert A_N^{-1} \Big( V_{N,int}^2 - V_N^2 (\gamma_0) \Big)   (A_N^{-1})^T \big\Vert_F >\f{\epsilon}{2}    \Big)
		\leq \P  \left(  \bar{M} \delta \cdot \big\Vert A_N^{-1} V_N^2 (\gamma_0)  (A_N^{-1})^T \big\Vert_F >\f{\epsilon}{2}   \right) \nn \\
		\leq &  \P  \left(  \bar{M} \delta \cdot ( \big\Vert \zeta^T\zeta \big\Vert_F +1 ) >\f{\epsilon}{2}   \right) + \P  \left(   \big\Vert \zeta^T\zeta \big\Vert_F +1 < \big\Vert A_N^{-1} V_N^2 (\gamma_0)   (A_N^{-1})^T \big\Vert_F  \right) \rightarrow 0, \label{half}
		\end{align}
		by let $\delta$ goes to $0$ and $N$ goes to $\infty$. On the other hand, similar to (\ref{Tn difference}), we have
		\begin{align*}
		&\big\Vert A_N^{-1} \Big(V_N^2 (\xi) - V_{N,int}^2 \Big)   (A_N^{-1})^T \big\Vert_F   \\
		\leq &  \Vert \Lambda^{-1/2} \Vert_F^2 \cdot   \Vert P \Vert_F^4  \cdot \max_{q\leq t\leq N}\bigg\vert\f{R_t(\xi)}{R_t(\gamma_0)} \bigg\vert \cdot 
		\bigg\Vert  \sum_{t=q }^N \bigg[   \left( \frac{\partial \log(\mu_t)  }{\partial \gamma} \right)^T  \left( \frac{\partial \log(\mu_t)  }{\partial \gamma} \right) \bigg|_{\gamma=\gamma_0} \\
		&\qquad \qquad \qquad \qquad \qquad \qquad \qquad  -\left( \frac{\partial \log(\mu_t)  }{\partial \gamma} \right)^T  \left( \frac{\partial \log(\mu_t)  }{\partial \gamma} \right) \bigg|_{\gamma=\xi}    \bigg] \cdot \mu_{t}|_{\gamma_0} \bigg\Vert_{F} \\
		\leq & (p+q)^3 \Big( \sum_{k=1}^{p+q} \f{1}{\lambda_k}   \Big) \sum_{t=q}^{N}  \phi^{[t/q]} \bar{M}' \cdot \sqrt{q} \Vert \hat{\gamma}-\gamma_0 \Vert_2 \cdot \mu_t\big|_{\gamma_0}
		\end{align*}
		for some constant $\bar{M}'$ depends on $\Vert \theta(\gamma_0) \Vert_1$ but not $t$ nor $N$. Thereby, same as (\ref{Tn difference prob}), we have 
		\begin{equation}
		\P\Big(   \big\Vert A_N^{-1} \Big(  V_N^2 (\xi)-V_{N,int}^2  \Big)   (A_N^{-1})^T \big\Vert_F >\f{\epsilon}{2}    \Big) \rightarrow 0, \label{other half}
		\end{equation}
		for which combine with (\ref{half}) would leads us to 
		\begin{equation}
		A_N^{-1}  \big( V_N^2 (\xi) - V_N^2 (\gamma_0) \big)  (A_N^{-1})^T \overset{P}{\longrightarrow} 0_{(p+q)\times (p+q)}. \label{2}
		\end{equation} 
		Besides, it's obvious that for $\forall \alpha \in \mathcal{M}_{(p+q)\times 1}$,
		\begin{align*}
		&\f{\alpha^T}{\Vert \alpha \Vert }A_N^{-1}T_{N,1}(\gamma_0)  (A_N^{-1})^T \f{\alpha}{\Vert \alpha \Vert } \\
		=&  -\sum_{t=q}^{N} \f{\alpha^T}{\Vert \alpha \Vert } A_N^{-1}   \frac{\partial \log(\mu_t)  }{\partial \gamma \partial \gamma^T} \Big|_{\gamma=\gamma_0}  (A_N^{-1})^T \f{\alpha}{\Vert \alpha \Vert } (I_t-\mu_t) \triangleq \sum_{t=q}^{N} \tilde{b}_t (I_t-\mu_t)
		\end{align*}
		is a martingale sequence, with
		\begin{align}
		|\tilde{b}_t| &\leq \f{\Vert \alpha \Vert_F^2}{\Vert \alpha \Vert_2^2}\cdot  \Vert \Lambda^{-1/2} \Vert_F^2 \cdot   \Vert P \Vert_F^4 \cdot \sqrt{2pq+q^2}  \max_{ \substack{1\leq k\leq (p+q),\\ 1\leq m\leq q} }  \Big\vert  \frac{\partial \log(\mu_t)  }{\partial \gamma_k \partial \theta_m } \big|_{\gamma_0}  \Big \vert \nn \\
		&\leq (p+q)^3 \Big( \sum_{k=1}^{p+q} \f{1}{\lambda_k}   \Big)   \phi^{[t/q]} \tilde{M}''. \label{tildeb}
		\end{align}
		Hence from (\ref{CM4}) of condition \ref{Condition for mixture normal},
		\begin{align*}
		\e \left[    \sum_{t=q}^{N} \tilde{b}_t (I_t-\mu_t)    \right]^2 &\leq   \sum_{t=q}^{N}  |\tilde{b}_t|^2 \cdot  \e \big( I_t-\mu_t   \big)^2 \\
		&\leq (p+q)^8 \tilde{M}''^2  \big(   \lambda_{\min}(A_N)  \big)^{-2} \sum_{t=q}^{N}  \phi^{[t/q]} \e \mu_t \rightarrow 0,
		\end{align*}
		and by Chebyshev's inequality that
		$
		(\alpha^T/\Vert \alpha \Vert ) A_N^{-1}T_{N,1}(\gamma_0) (A_N^{-1})^T (\alpha/\Vert \alpha \Vert) \rightarrow_p 0,
		$
		and since $\alpha$ is arbitrary, so it's equivalent to say
		\begin{equation}
		A_N^{-1} T_{N,1}(\gamma_0) (A_N^{-1})^T \overset{P}{\longrightarrow} 0_{(p+q)\times (p+q)}. \label{4}
		\end{equation}
		For the last term which relates to $T_{N,2}(\xi)$, we introduce an intermediate $T_{N,int}$ defined as 
		\begin{equation*}
		T_{N,int}=\sum_{t=q}^{N}  \frac{\partial \log(\mu_t)  }{\partial \gamma \partial \gamma^T} \Big|_{\gamma=\gamma_0} (\mu_t|_{\xi}-\mu_t|_{\gamma_0}),
		\end{equation*}
		then similar to (\ref{Tn difference}), we may derive
		\begin{align*}
		&\big\Vert A_N^{-1} \Big(T_{N,2} (\xi) - T_{N,int} \Big)   (A_N^{-1})^T \big\Vert_F   \\
		\leq &  \Vert \Lambda^{-1/2} \Vert_F^2 \cdot   \Vert P \Vert_F^4  \cdot \max_{q\leq t\leq N} \bigg\vert\f{R_t(\xi)}{R_t(\gamma_0)} -1 \bigg\vert \cdot 
		\bigg\Vert  \sum_{t=q }^N \bigg[   \frac{\partial \log(\mu_t)  }{\partial \gamma \partial \gamma^T}  \bigg|_{\gamma=\gamma_0} - \frac{\partial \log(\mu_t)  }{\partial \gamma \partial \gamma^T }\bigg|_{\gamma=\xi}    \bigg] \cdot \mu_{t}|_{\gamma_0} \bigg\Vert_{F} \\
		\leq & (p+q)^3 \Big( \sum_{k=1}^{p+q} \f{1}{\lambda_k}   \Big) \sum_{t=q}^{N}  \phi^{[t/q]} \bar{M}' \cdot \sqrt{q} \Vert \hat{\gamma}-\gamma_0 \Vert_2^2 \cdot \mu_t\big|_{\gamma_0}
		\end{align*}
		so same as (\ref{Tn difference prob}), we conclude 
		\begin{equation}
		A_N^{-1}  \Big(T_{N,2} (\xi) - T_{N,int} \Big)  (A_N^{-1})^T \overset{P}{\longrightarrow} 0_{(p+q)\times (p+q)}. \label{5}
		\end{equation} 
		while for $A_N^{-1}  T_{N,int} (A_N^{-1})^T$, we again consider its one-dimensional counterpart. For $\forall \alpha \in \mathcal{M}_{(p+q)\times 1}$, and under the event $\{ \Vert \hat{\gamma} - \gamma_0 \Vert_2 \leq \delta\}$ for some $\delta\leq 1$, we have
		\begin{equation*}
		\bigg\vert \f{\alpha^T}{\Vert \alpha \Vert }A_N^{-1}T_{N,int} (A_N^{-1})^T \f{\alpha}{\Vert \alpha \Vert } \bigg\vert = \bigg\vert \sum_{t=q}^{N} \tilde{b}_t \bigg[ \f{R_t(\xi)}{R_t(\gamma_0)} -1 \bigg] \mu_t |_{\gamma_0} \bigg\vert \leq \bar{M} \delta  \sum_{t=q}^{N}  \big\vert \tilde{b}_t \big\vert \cdot  \mu_t |_{\gamma_0}.
		\end{equation*}
		Using (\ref{tildeb}) and we obtain
		\begin{align*}
		&\P \left(    \bigg\vert \f{\alpha^T}{\Vert \alpha \Vert }A_N^{-1}T_{N,int} (A_N^{-1})^T \f{\alpha}{\Vert \alpha \Vert } \bigg\vert  >\epsilon    \right) \\
		\leq &\P \Big(  \Vert \hat{\gamma} - \gamma_0 \Vert_2 > \delta   \Big) + \P \left(  \bar{M} \delta  \sum_{t=q}^{N}  \big\vert \tilde{b}_t \big\vert \cdot  \mu_t |_{\gamma_0} >\epsilon  \right) \\
		\leq &\P \Big(  \Vert \hat{\gamma} - \gamma_0 \Vert_2 > \delta   \Big) + \f{\bar{M} \delta  \sum_{t=q}^{N}  |\tilde{b}_t| \cdot  \e \mu_t }{\epsilon} \\ 
		\leq &\P \Big(  \Vert \hat{\gamma} - \gamma_0 \Vert_2 > \delta   \Big) + \f{\bar{M} \delta   }{\epsilon}\cdot (p+q)^4 \tilde{M}''  \big(   \lambda_{\min}(A_N)  \big)^{-1} \sum_{t=q}^{N}  \phi^{[t/q]} \e \mu_t \rightarrow 0
		\end{align*}
		by let $\delta$ goes to $0$ and $N$ goes to $\infty$. Also, since $\alpha$ is arbitrary, so it's equivalent to say that 
		\begin{equation}
		A_N^{-1}  T_{N,int} (A_N^{-1})^T \overset{P}{\longrightarrow} 0_{(p+q)\times (p+q)}. \label{6}
		\end{equation}
		
		Overall, combine (\ref{1}), (\ref{2}) and (\ref{4})-(\ref{6}), and the (\ref{CM1}) of condition \ref{Condition for mixture normal} that $\tilde{V}_N^2= A_N^{-1} V_N^2(\gamma_0) (A_N^{-1})^T \overset{P}{\longrightarrow} \zeta^T\zeta$, implies
		\begin{equation}
		A_N^{-1}  \left( - \f{\partial U_N(\gamma)}{\partial \gamma}\Big|_{\gamma=\xi }   \right) (A_N^{-1})^T  \overset{P}{\longrightarrow}  \zeta^T\zeta .  \label{weight}
		\end{equation}
		Thus, by (\ref{distribution}) and (\ref{weight}), we conclude that,
		\begin{equation}
		\zeta A_N^T \big(  \hat{\gamma} -\gamma_0    \big) \overset{P}{\longrightarrow} Z \label{main result}
		\end{equation}
		on the non-extinction set $\mathcal{E}_{none}$ with $\zeta$ being independent of $Z\sim \mathcal{N}(0,I)$, and since we have chosen $\zeta$ and $A_N$ to be symmetric, so we may also choose a symmetric matrix
		\begin{equation}
		\zeta A_N^T \Big[ \sum_{t=q}^{N}\cov  \big(   \xi_t(\gamma_0) | \mathcal{F}_{t-1}    \big)	  \Big]^{-1/2} \rightarrow I   \label{va}
		\end{equation}
		thus we may completes the proof and obtain (3.1) by plug (\ref{va}) into (\ref{main result}). Last, one thing worth mention is, the $\tilde{V}_N(\alpha)-1=\sum b_t (I_t-\mu_t)$ can be written into a martingale difference array in this very special case, so (\ref{CM1}) of condition \ref{condition quasi} could be replaced by some other ``martingale array convergence condition". However, only little work has been done towards this end and seems like neither the condition in \cite{ghosal1998complete} (for complete convergence) or the one in \cite{atchade2009strong} is easy tracking, but it would certainly be interesting to see if one could reduce (\ref{CM1}) of condition \ref{condition quasi}.
	\end{proof}

\section*{C. Proofs of theorems in Section 3.2}
	\setcounter{equation}{0}
	\renewcommand{\theequation}{C.\arabic{equation}}
	
	\begin{proof} (\emph{of Theorem 3.2})
		\noindent Since $\tilde{\beta}^{(k)}$, as well as $\log \tilde{R}_t^{(k)}$, $t=\tau_0+1,\cdots,k$ are linear functions of $\theta$, so there exists constants $a_t, b_t \in \mathcal{M}_{q\times 1}$ depend on $\mathcal{F}_{k-1}$ such that
		\begin{equation*}
		\log \tilde{R}_t^{(k)} = a_t +b_t^T \theta, \quad {\rm for}\;\; t=\tau_0+1,\cdots,k.
		\end{equation*}
		Consequently, by plug it into the right hand side of (3.3), we obtain the profile likelihood
		\begin{equation*}
		\tilde{\ell}_{k}= \sum_{t=\tau_0+1}^{k}  I_{t} (a_t +b_t^T \theta) - e^{(a_t +b_t^T \theta)} \Lambda_{t},
		\end{equation*}
		and the Hessian matrix of $\tilde{\ell}_{k}$ with respect to $\theta$ is
		\begin{align*}
		H_{k} &= \f{\partial^2 \tilde{\ell}_{k} }{ \partial \theta \partial \theta^T} 
		=  - \sum_{t=\tau_0+1}^{k} \tilde{R}_t^{(k)} \Lambda_{t} 
		b_t b_t^T \leq 0. 
		\end{align*}
		Hence (3.3) is a global concave maximization problem. 
	\end{proof}
	
	\begin{proof} (\emph{of Theorem \ref{thmconcave2}})
		\noindent Due to the fact that $\tilde{R}_{t}^{(k)}$ and $\tilde{g}_{h,k}$ are linear functions of $\phi_0$ and $\theta$, so it follows the same manner of the proof of theorem 3.2. There exists constants $(a_t,b_t^T)$ depend on $\mathcal{F}_{t-1}$ such that
		\begin{equation*}
		\tilde{R}_t^{(k)} = a_t +b_t^T \theta, \quad {\rm for}\;\; t=\tau_0+1,\cdots,k.
		\end{equation*}
		Consequently, the likelihood and its Hessian matrix are
		\begin{align*}
		\tilde{\ell}_{k}&= \sum_{t=\tau_0+1}^{k}  I_{t} \log \big( a_t +b_t^T \theta \big) -  (a_t +b_t^T \theta) \Lambda_{j}, \\
		H_{k}&= - \sum_{t=\tau_0+1}^{k} \f{I_t}{\big( a_t +b_t^T \theta  \big)^2} 
		b_tb_t^T\leq 0. 
		\end{align*}
		Hence, Equation (\ref{nonpara}) is a global concave maximization problem.  
	\end{proof}
	
	\begin{condition}	\label{condition2}
		When given $\{  (I_t,Z_t) \}_{1\leq t\leq k}  $, consider the backward looking estimator of $\gamma$, for $\varphi_{t} \triangleq \log(\tilde{R}_{t})-\theta_q$, there exist constants $C_{\varphi}$, $C_{d\varphi}$, $C_{v\varphi}$, such that, for $1\leq m \leq q-1$,
		\begin{gather*}
		\max_{1\leq t \leq k} | \varphi_t |   \leq C_{\varphi}, \quad 	\max_{1\leq t \leq k} \Big|\f{ \partial  \varphi_t  }{\partial \theta_m} \Big|   \leq C_{d\varphi}, \quad
		\f{1}{k^2}\sum_{\tau_0+1\leq t<s \leq k}  \Big(  \f{ \partial  \varphi_t  }{\partial \theta_m}  -\f{ \partial  \varphi_s  }{\partial \theta_m}    \Big)^2 \geq C_{v\varphi}.
		\end{gather*}
	\end{condition}
	\begin{remark}
		Condition \ref{condition2} is a restriction on $\{ Z_t \}_{t\geq 1}$ and it's loose consider the high variability this series of covariates have. In general, the condition is hard to verify despite been easily satisfied, an strict and sufficient alternative for the first two condition is to require $\max_{k\geq \tau_0} \sum_{t=1}^{k} | Z_{k+1}^T H_k^{-1}(Z_t-\bar{Z}_k)| \leq (1-\Vert \theta \Vert_1)/2 $, where $H_k=\sum_{t=1}^{k} (Z_t-\bar{Z}_k)(Z_t-\bar{Z}_k)^T$.
	\end{remark}

	\begin{proof}(\emph{of Theorem 3.3})
		We first eliminate the trivial cases where the pandemic cases vanishes or diverge to infinite, the first indicates the pandemic ends and the second is by nature unreasonable due to herd immunity or population limits. Thereby, without those two trivial cases, when given $\{ (I_t,Z_t)_{1\leq t \leq k} \}$, we have $\max\Lambda_t$ is bounded from above by $C_{u\lambda}$ and bounded away from zero by $C_{\lambda}$. Now, notice that $\tilde{\beta}^{(k)}$ is essentially just a linear function of $\theta$ and does not involve $\phi_0$, thereby, a simple reorganization of (3.3) leads to
		\begin{align*}
		\tilde{\ell}_{k} 
		&\triangleq  \sum_{t=\tau_0+1}^{k}  \Big(  I_t \log \tilde{R}_t^{(k)} - \tilde{R}_t^{(k)} \Lambda_{t} \Big) \\
		& = \theta_q \big( \sum_{t=\tau_0+1}^{k}  I_{t} \big) -  e^{\theta_q} \big( \sum_{t=\tau_0+1}^{k}  e^{\sum_{m=1}^{q-1} \theta_m \log(\hat{R}_{t-m}^{(k-1)}) + Z_{t}^T \tilde{\beta}^{(k)} } \Lambda_{t} \big) \\
		&\quad + \sum_{t=\tau_0+1}^{k} I_t \Big( \sum_{m=1}^{q-1} \theta_m \log(\hat{R}_{t-m}^{(k-1)}) + Z_{t}^T \tilde{\beta}^{(k)}\Big),
		\end{align*}
		Notice that
		\begin{equation*}
		\f{\partial^2 \tilde{\ell}_{k} }{ \big( \partial \theta_q \big)^2 } =- e^{\theta_q} \big( \sum_{t=\tau_0+1}^{k}  e^{\sum_{m=1}^{q-1} \theta_m \log(\hat{R}_{t-m}^{(k-1)}) + Z_{t}^T \tilde{\beta}^{(k)} } \Lambda_{t} \big)  \leq 0,
		\end{equation*}
		so by letting $\partial \tilde{\ell}_{k} / \partial \theta_q =0$, we have $ \hat{\theta}^{(k)}$ satisfy the equations
		\begin{align}
		\exp\big( \hat{ \theta}_q^{(k)}  \big) &=  \big( \sum_{t=\tau_0+1}^{k}  I_{t} \big) \cdot \Big( \sum_{t=\tau_0+1}^{k}  e^{\sum_{m=1}^{q-1} \theta_m \log(\hat{R}_{t-m}^{(k-1)}) + Z_{t}^T \tilde{\beta}^{(k)} } \Lambda_{t} \Big)^{-1} , \nn \\
		\hat{\theta}^{(k)} &=  \arg\max_{\Vert \theta \Vert_1 <1} \bigg[  \sum_{t=\tau_0+1}^{k} I_t \Big( \sum_{m=1}^q \theta_m \log(\hat{R}_{t-m}^{(k-1)}) + Z_{t}^T \tilde{\beta}^{(k)}\Big)  \nn  \\
		&\quad - \big( \sum_{t=\tau_0+1}^{k}  I_{t} \big) \log \Big( \sum_{t=\tau_0+1}^{k}  e^{\sum_{m=1}^q \theta_m \log(\hat{R}_{t-m}^{(k-1)}) + Z_{t}^T \tilde{\beta}^{(k)} } \Lambda_{t} \Big)  \bigg] \triangleq \arg\max_{\Vert \theta \Vert_1 <1}  \tilde{\ell}_{\theta,k}, \label{equationphi}  
		\end{align}
		or equivalently, plug the form of optimal $\phi_0$ into the quasi-score estimating equation and conclude $\hat{\theta}^{(k)}$ satisfy $\partial  \tilde{\ell}_{\theta,k} / \partial \theta =0 $, i.e.,
		\begin{align}
		0= - \big( &\sum_{t=\tau_0+1}^{k}  I_{t} \big)   \f{  \sum_{t=\tau_0+1}^{k}  e^{ \varphi_{t}^{(k)} } \Lambda_{t}  \f{ \partial  \varphi_{t}^{(k)} }{ \partial \theta } }{    \sum_{t=\tau_0+1}^{k}  e^{ \varphi_{t}^{(k)} } \Lambda_{t}    }\bigg|_{\theta=\hat{\theta}^{(k)}}  + \sum_{t=\tau_0+1}^{k}  I_t  \f{ \partial  \varphi_{t}^{(k)} }{ \partial \theta } , \label{iterative mapping}
		\end{align}
		where $ \varphi_{t}^{(k)}= \sum_{m=1}^{q-1} \theta_m \log(\hat{R}_{t-m}^{(k-1)}) + Z_{t}^T \tilde{\beta}^{(k)} $, and for $1\leq m\leq q$,
		\begin{gather*}
		\f{ \partial \varphi_{t}^{(k)} }{ \partial \theta_m } =  \log \big( \hat{R}_{t-m}^{(k-1)} \big) + Z_{t}^T \f{ \partial \tilde{\beta}^{(k)} }{ \partial \theta_m} , \quad  \f{ \partial^2 \varphi_{t}^{(k)} }{ \partial \theta \partial \theta^T } =0.
		\end{gather*}
		Meanwhile, Cauchy's inequality shows that for each $k\geq \tau_0+1$,
		\begin{align}
		&\Big(  \sum_{t=\tau_0+1}^{k}  e^{\varphi_{t}^{(k)}} \Lambda_{t}  \Big)^2 \f{ \partial^2  \tilde{\ell}_{\theta,k} }{ \partial \theta \partial \theta^T } \bigg/  \big( \sum_{t=\tau_0+1}^{k}  I_{t} \big) \nn \\
		=& - \bigg[ \sum_{t=\tau_0+1}^{k}  e^{\varphi_{t}^{(k)}} \Lambda_{t} \bigg( \f{ \partial \varphi_{t}^{(k)}}{ \partial \theta } \bigg) \bigg( \f{ \partial \varphi_{t}^{(k)}}{ \partial \theta } \bigg)^T \bigg]\Big(  \sum_{t=\tau_0+1}^{k}  e^{\varphi_{t}^{(k)}} \Lambda_{t}  \Big) \nn \\
		&\qquad + \bigg[ \sum_{t=\tau_0+1}^{k}  e^{\varphi_{t}^{(k)}} \Lambda_{t}  \bigg( \f{ \partial \varphi_{t}^{(k)} }{ \partial \theta } \bigg) \bigg]\bigg[ \sum_{t=\tau_0+1}^{k}  e^{\varphi_{t}^{(k)}} \Lambda_{t}  \bigg( \f{ \partial \varphi_{t}^{(k)} }{ \partial \theta } \bigg) \bigg]^T \leq 0,   \label{second derivative}	
		\end{align}
		thereby the solution of 
		%(\ref{equationphi}) 
		\begin{equation}
		\hat{\theta}^{(k)} \triangleq \arg\max_{\Vert \theta \Vert_1 < 1 }  \tilde{\ell}_{\theta,k} ,\label{equationphi2}
		\end{equation}
		would be a unique point or a closed interval. The second case happens only when the equality sign of (\ref{second derivative}) holds, i.e., 
		\begin{equation*}
		\f{ \partial \varphi_{t}^{(k)}}{ \partial \theta }  = C_0, \quad t=\tau_0+1,\cdots,k,
		\end{equation*}
		for some constant $C_0$. This implies the quasi-score equation (\ref{iterative mapping}) holds for all $\theta \in \{ \Vert \theta \Vert_1<1 \}$, for instance the case where $\theta=0$, which is trivial since (2.3) degenerate to linear regression and we omit discussions on this scenario.    
		On the contrast, we have $\hat{\theta}^{(k)}$ is the unique solution of (\ref{iterative mapping}) and $\hat{\theta}^{(k-1)}$ satisfy the same equation with $k$ replaced by $k-1$, i.e,	
		\begin{align}
		0= &\sum_{t=\tau_0+1}^{k-1}  I_t  \f{ \partial  \varphi_{t}^{(k-1)} }{ \partial \theta } - \big( \sum_{t=\tau_0+1}^{k-1}  I_{t} \big)  \f{  \sum_{t=\tau_0+1}^{k-1}  e^{ \varphi_{t}^{(k-1)} } \Lambda_{t}  \f{ \partial  \varphi_{t}^{(k-1)} }{ \partial \theta } }{    \sum_{t=\tau_0+1}^{k-1}  e^{ \varphi_{t}^{(k-1)}  } \Lambda_{t}    } . \label{iterative mapping 2}
		\end{align}
		Combine (\ref{iterative mapping}) and (\ref{iterative mapping 2}) leads to 
		\begin{align*}
		&\big( \sum_{t=\tau_0+1}^{k}  I_{t} \big)   \f{  \sum_{t=\tau_0+1}^{k}   e^{ \varphi_{t}^{(k)} } \Lambda_{t}  \f{ \partial  \varphi_{t}^{(k)} }{ \partial \theta } }{    \sum_{t=\tau_0+1}^{k}   e^{ \varphi_{t}^{(k)} } \Lambda_{t}  }\bigg|_{\theta=\hat{\theta}^{(k)}}= \sum_{t=\tau_0+1}^{k}  I_t   \f{ \partial  \varphi_{t}^{(k)} }{ \partial \theta } \\
		=&  \big( \sum_{t=\tau_0+1}^{k-1}  I_{t} \big)   \f{  \sum_{t=\tau_0+1}^{k-1}   e^{ \varphi_{t}^{(k-1)} } \Lambda_{t}  \f{ \partial  \varphi_{t}^{(k-1)} }{ \partial \theta } }{    \sum_{t=\tau_0+1}^{k-1}   e^{ \varphi_{t}^{(k-1)} } \Lambda_{t}  }\bigg|_{\theta=\hat{\theta}^{(k-1)}} + I_{k} \f{ \partial  \varphi_{k}^{(k)} }{ \partial \theta },
		\end{align*}
		hence
		\begin{align*}
		&\big( \sum_{t=\tau_0+1}^{k}  I_{t} \big)   \f{  \sum_{t=\tau_0+1}^{k}   e^{ \varphi_{t}^{(k)} } \Lambda_{t}  \f{ \partial  \varphi_{t}^{(k)} }{ \partial \theta } }{    \sum_{t=\tau_0+1}^{k}   e^{ \varphi_{t}^{(k)} } \Lambda_{t}  }\bigg|_{\theta=\hat{\theta}^{(k)}}\\
		-&\big( \sum_{t=\tau_0+1}^{k-1}  I_{t} \big)   \f{  \sum_{t=\tau_0+1}^{k-1}   e^{ \varphi_{t}^{(k-1)} } \Lambda_{t}  \f{ \partial  \varphi_{t}^{(k-1)} }{ \partial \theta } }{    \sum_{t=\tau_0+1}^{k-1}   e^{ \varphi_{t}^{(k-1)} } \Lambda_{t}  }\bigg|_{\theta=\hat{\theta}^{(k)}} -  I_{k} \f{ \partial  \varphi_{k}^{(k)} }{ \partial \theta }\\
		=&  \big( \sum_{t=\tau_0+1}^{k-1}  I_{t} \big)   \f{  \sum_{t=\tau_0+1}^{k-1}   e^{ \varphi_{t}^{(k-1)} } \Lambda_{t}  \f{ \partial  \varphi_{t}^{(k-1)} }{ \partial \theta } }{    \sum_{t=\tau_0+1}^{k-1}   e^{ \varphi_{t}^{(k-1)} } \Lambda_{t}  }\bigg|_{\theta=\hat{\theta}^{(k-1)}}-  \big( \sum_{t=\tau_0+1}^{k-1}  I_{t} \big)   \f{  \sum_{t=\tau_0+1}^{k-1}   e^{ \varphi_{t}^{(k-1)} } \Lambda_{t}  \f{ \partial  \varphi_{t}^{(k-1)} }{ \partial \theta } }{    \sum_{t=\tau_0+1}^{k-1}   e^{ \varphi_{t}^{(k-1)} } \Lambda_{t}  }\bigg|_{\theta=\hat{\theta}^{(k)}} ,
		\end{align*}
		and it further implies
		\begin{align}
		&\f{ \partial^2  \tilde{\ell}_{\theta,k-1} }{ \partial \theta \partial \theta^T } \bigg|_{\theta=\theta^*} \cdot  (\hat{\theta}^{(k)}- \hat{\theta}^{(k-1)}) \nn \\
		= & \big( \sum_{t=\tau_0+1}^{k-1}  I_{t} \big)\bigg[   \f{  e^{\varphi_{k}^{(k)} } \Lambda_{k}  \f{ \partial \varphi_{k}^{(k)} }{ \partial \theta }+ \sum_{t=\tau_0+1}^{k-1}  e^{\varphi_{t}^{(k)}} \Lambda_{t}  \f{ \partial \varphi_{t}^{(k)} }{ \partial \theta } }{   e^{\varphi_{k}^{(k)} } \Lambda_{k} +  \sum_{t=\tau_0+1}^{k-1}  e^{\varphi_{t}^{(k)} } \Lambda_{t}    } -\f{  \sum_{t=\tau_0+1}^{k-1}  e^{\varphi_{t}^{(k-1)}} \Lambda_{t}  \f{ \partial \varphi_{t}^{(k-1)} }{ \partial \theta } }{    \sum_{t=\tau_0+1}^{k-1}  e^{\varphi_{t}^{(k-1)}} \Lambda_{k-1}    } \bigg] \bigg|_{\theta=\hat{\theta}^{(k)}} \nn \\
		& +  I_{k} \bigg[   \f{  \sum_{t=\tau_0+1}^{k}  e^{\varphi_{t}^{(k)}} \Lambda_{t}  \f{ \partial \varphi_{t}^{(k)} }{ \partial \theta } }{    \sum_{t=\tau_0+1}^{k}  e^{\varphi_{t}^{(k)}} \Lambda_{t}    }\bigg|_{\theta=\hat{\theta}^{(k)}} - \f{ \partial \varphi_{k}^{(k)} }{ \partial \theta } \bigg] \label{iteration equation}
		\end{align}
		for some $\theta^*$ between $\hat{\theta}^{(k-1)}$ and $\hat{\theta}^{(k)}$. Since $\psi_t$ is actually invariant over $k\geq t$, so
		\begin{align}
		&\bigg[   \f{ \sum_{t=\tau_0+1}^{k}  e^{\varphi_{t}^{(k)}} \Lambda_{t}  \f{ \partial \varphi_{t}^{(k)} }{ \partial \theta } }{   \sum_{t=\tau_0+1}^{k}  e^{\varphi_{t}^{(k)} } \Lambda_{t}    } -\f{  \sum_{t=\tau_0+1}^{k-1}  e^{\varphi_{t}^{(k-1)}} \Lambda_{t}  \f{ \partial \varphi_{t}^{(k-1)} }{ \partial \theta } }{    \sum_{t=\tau_0+1}^{k-1}  e^{\varphi_{t}^{(k-1)}} \Lambda_{k-1}    } \bigg] \bigg|_{\theta=\hat{\theta}^{(k)}} \nn \\
		=&    \f{  \sum_{t=\tau_0+1}^{k-1} e^{\varphi_{k}} \Lambda_{k} e^{\varphi_{t}} \Lambda_{t}  \big(  \f{ \partial \varphi_k} { \partial \theta }  -\f{ \partial \varphi_t} { \partial \theta }\big)  }{   \big( \sum_{t=\tau_0+1}^{k}  e^{\varphi_{t}^{(k)}     } \Lambda_{t}   \big) \big( \sum_{s=\tau_0+1}^{k-1}  e^{\varphi_{s}^{(k-1)} } \Lambda_{s}   \big)   } \bigg|_{\theta=\hat{\theta}^{(k)}} = O(\f{1}{k}) \label{thm2eq1}
		\end{align}
		By condition \ref{condition2}, and for each $1\leq m\leq q-1$, we have
		\begin{align*}
		&\bigg| \f{ \partial^2  \tilde{\ell}_{\theta,k-1} }{( \partial \theta_m)^2 } \bigg|_{\theta=\theta^*} \big/ \big( \sum_{t=\tau_0+1}^{k-1}  I_{t} \big)   \\
		=& \f{ \bigg( \sum_{t=\tau_0+1}^{k-1}  e^{ \varphi_{t}^{(k-1)} } \Lambda_{t} \bigg( \f{ \partial \varphi_{t}^{(k-1)} }{ \partial \theta_m } \bigg)^2 \bigg)\Big(  \sum_{t=\tau_0+1}^{k-1}  e^{ \varphi_{t}^{(k-1)} } \Lambda_{t}   \Big) }{   \Big(  \sum_{t=\tau_0+1}^{k-1}  e^{\varphi_{t}^{(k-1)} } \Lambda_{t} \Big)^2   } -  \f{ \bigg( \sum_{t=\tau_0+1}^{k-1}  e^{\varphi_t^{(k-1)}} \Lambda_{t}  \bigg( \f{ \partial \varphi_t^{(k-1)} }{ \partial \theta_m } \bigg)  \bigg)^2 }{  \Big(  \sum_{t=\tau_0+1}^{k-1}  e^{\varphi_{t}^{(k-1)} } \Lambda_{t} \Big)^2  } \\
		=& \f{ \sum_{\tau_0+1 \leq t<s \leq k-1}  e^{\varphi_t^{(k-1)}+\varphi_s^{(k-1)} } \Lambda_{t} \Lambda_s  \bigg( \f{ \partial \varphi_t^{(k-1)} }{ \partial \theta_m } -  \f{ \partial \varphi_s^{(k-1)} }{ \partial \theta_m } \bigg)^2 }{   \Big(  \sum_{t=\tau_0+1}^{k-1}  e^{\varphi_{t}^{(k-1)} } \Lambda_{t} \Big)^2   } \geq e^{-4C_{\varphi}} \f{C^2_{\lambda}}{C^2_{u\lambda}} C_{v\varphi} .
		%\geq & \sum_{1\leq j<k \leq i-\tau_0-1} e^{-2B_1}C_{\lambda 1}^2  \big( \gamma_{(j)}-\gamma_{(k)}  \big)^2 \\
		%\geq& e^{-2B_1}C_{\lambda 1}^2  \sum_{j=1}^{C_{\lambda 4} i} \sum_{k=j+1}^{C_{\lambda 4} i}  \f{C_{\lambda 3}^2 (k-j)^2 }{i^2} \asymp \f{C_{\lambda 3}^2 C_{\lambda 4}^4 i^2}{12}.
		\end{align*}
		\noindent Combine with (\ref{iteration equation}) and (\ref{thm2eq1}) implies that
		%, $\exists$ constant $C$, s.t.,
		%	\begin{align*}
		%	 C \cdot \big| \hat{\theta}_{s}^{(i+1)}- \hat{\theta}_s^{(i)} \big|  &\leq \f{2e^{4C_\varphi}C_{u\lambda}^2C_{d\varphi} I_{i+1} }{C_{\lambda}^2\big( \sum_{j=\tau_0+1}^{i}  I_{j} \big)} + \f{e^{4C_\varphi}C_{u\lambda}^2 }{C_{\lambda}^2} \bigg[   \f{  e^{\varphi_{i+1}^{(i+1)} } \Lambda_{i+1} | \f{ \partial \varphi_{i+1}^{(i+1)} }{ \partial \theta_s } | }{   e^{\varphi_{i+1}^{(i+1)} } \Lambda_{i+1} +  \sum_{j=\tau_0+1}^{i}  e^{\varphi_{j}^{(i)}} \Lambda_{j}    } \\
		%	& \qquad + \f{  \big( \sum_{j=\tau_0+1}^{i}  e^{\varphi_{i+1}^{(i+1)}} \Lambda_{j} | \f{ \partial \varphi_{j}^{(i)} }{ \partial \theta_s } | \big) e^{\varphi_{i+1}^{(i+1)} } \Lambda_{i+1} }{  \big(  \sum_{j=\tau_0+1}^{i}  e^{\varphi_{j}^{(i)}} \Lambda_{j}  \big) \big(    e^{\varphi_{i+1}^{(i+1)} } \Lambda_{i+1} +  \sum_{j=\tau_0+1}^{i}  e^{\varphi_{j}^{(i)}} \Lambda_{j}  \big)    } \bigg]  \\
		%	&\leq \f{2e^{4C_\varphi}C_{u\lambda}^2C_{d\varphi} }{C_{\lambda}^2}  \f{  e^{\varphi_{i+1}^{(i+1)}} \Lambda_{i+1} }{  \sum_{j=\tau_0+1}^{i+1}  e^{\varphi_j^{(i)}} \Lambda_{j}    }+ \f{2e^{4C_\varphi}C_{u\lambda}^2C_{d\varphi} I_{i+1} }{C_{\lambda}^2\big( \sum_{j=\tau_0+1}^{i}  I_{j} \big)} ,
		%	\end{align*}
		%	which leads to 	  
		\begin{gather*}
		\big| \hat{\theta}_{m}^{(k)}- \hat{\theta}_m^{(k-1)} \big| = O\Big(   \f{1 }{k}+\f{I_{k}}{\big( \sum_{t=\tau_0+1}^{k-1}  I_{t} \big)}     \Big). 
		\end{gather*}
	\end{proof}

	When consider the forward-looking estimator, it behaves much more complicated, first, a stronger condition would be needed just to conclude a similar result, for instance, 
	
	\begin{condition}	\label{condition3}
		Suppose $\Vert \theta \Vert_1 <\phi <1$, $\max\Vert Z_t \Vert_1\leq M$, and when given $\{  (I_t,Z_t) \}_{1\leq t\leq k}  $, for $\varphi_{t}^{(k)} \triangleq \log(\tilde{R}_{t}^{(k)})-\theta_q$, there exist constants $C_{\varphi}$, $C_{d\varphi}$, $C_{v\varphi}$, s.t., for $1\leq m \leq q-1$, and each $k$,
		\begin{gather*}
		\max_{1\leq t \leq k} | \varphi_t^{(k)} |   \leq C_{\varphi}, \quad 	\max_{1\leq t \leq k} \Big|\f{ \partial  \varphi_t^{(k)}  }{\partial \theta_m} \Big|   \leq C_{d\varphi}, \quad
		\f{1}{k^2}\sum_{\tau_0+1\leq t<s \leq k}  \Big(  \f{ \partial  \varphi_t^{(k)}  }{\partial \theta_m}  -\f{ \partial  \varphi_s^{(k)}  }{\partial \theta_m}    \Big)^2 \geq C_{v\varphi}.
		\end{gather*}
	\end{condition}
	Second, the proof of Theorem 3.3 needs justification from line to line. Since now $\varphi_t^{(k)}$ is a term involve $k$, so we define $d_{\varphi_t}^{(k)}=| \hat{\varphi}_t^{(k)} - \hat{\varphi}_t^{(k-1)}|$ and similarly define $d_{\theta}^{(k)}= \Vert \theta^{(k)} -\theta^{(k-1)}  \Vert_1$. Since $\varphi_t^{(k)}$ is bounded uniformly over $k$, so we have $|e^{\varphi_t^{(k)}}-e^{\varphi_t^{(k)}}| \leq O (d_{\varphi_t}^{(k)})$.  
	%similarly define $d_{\tilde{\beta}}^{(k)}$, $d_{\hat{\beta}}^{(k)}$, $d_{\varphi_t}^{(k)}$, define
	% \begin{equation*}
	%        d_{R_t}^{(k)}=| \log \hat{R}_t^{(k)} - \log \hat{R}_t^{(k-1)}| ,\;\; d_{\theta}^{(k)}= \Vert \hat{\theta}^{(k)} - \hat{\theta}^{(k-1)} \Vert_1.
	% \end{equation*}
	% So apparently, 
	% \begin{align*}
	%        d_{\varphi_t}^{(k)}&\leq \sum_{s=1}^{q} \theta_s d_{R_{t-s}}^{(k-1)} + M d_{\tilde{\beta}}^{(k)} \leq \phi \max_{t-q\leq i\leq t-1} d_{R_{i}}^{(k-1)} + M d_{\tilde{\beta}}^{(k)}, \\
	%        d_{\phi}^{(k)} &\leq \log \left\{  \left[ 1+ \f{I_k}{\sum_{t=\tau_0+1}^{k-1} I_t}  \right] \left[  1+ O\bigg(  \max_t \exp (d_{\varphi_t}^{(k)})   \bigg) \right]     \right\}\\
	%        &=\f{I_k}{\sum_{t=\tau_0+1}^{k-1} I_t} + O\bigg(  \max_t \exp (d_{\varphi_t}^{(k)})   \bigg) , \\
	%        d_{R_t}^{(k)} & \leq d_{\phi}^{(k)}+ Md_{\hat{\beta}}^{(k)} + \sum_{s=1}^{q} \theta_s d_{R_{t-s}}^{(k-1)} \leq d_{\phi}^{(k)}+ Md_{\hat{\beta}}^{(k)} +  \phi \max_{t-q\leq i\leq t-1} d_{R_{i}}^{(k-1)} 
	% \end{align*}
	% Therefore, 
	Consequently, (\ref{thm2eq1}) for forward-looking estimator should be justified to 
	\begin{align*}
	&\bigg[   \f{ \sum_{t=\tau_0+1}^{k}  e^{\varphi_{t}^{(k)}} \Lambda_{t}  \f{ \partial \varphi_{t}^{(k)} }{ \partial \theta } }{   \sum_{t=\tau_0+1}^{k}  e^{\varphi_{t}^{(k)} } \Lambda_{t}    } -\f{  \sum_{t=\tau_0+1}^{k-1}  e^{\varphi_{t}^{(k-1)}} \Lambda_{t}  \f{ \partial \varphi_{t}^{(k-1)} }{ \partial \theta } }{    \sum_{t=\tau_0+1}^{k-1}  e^{\varphi_{t}^{(k-1)}} \Lambda_{k-1}    } \bigg] \bigg|_{\theta=\hat{\theta}^{(k)}} \nn \\
	\leq &  O(\f{1}{k})+ \max_t \f{\partial d_{\varphi_t}^{(k)}}{\partial \theta}+  \f{  \sum_{t=\tau_0+1}^{k-1}  \sum_{s=\tau_0+1}^{k-1}  \f{ \partial \varphi_{t}^{(k-1)} }{ \partial \theta }\Lambda_{t}  \Lambda_{s} \big(  \exp ( \varphi_t^{(k)}+\varphi_{s}^{(k-1)} )  -  \exp( \varphi_t^{(k-1)}+\varphi_{s}^{(k)}  )  \big)  }{   \big( \sum_{t=\tau_0+1}^{k-1}  e^{\varphi_{t}^{(k)}     } \Lambda_{t}   \big) \big( \sum_{s=\tau_0+1}^{k-1}  e^{\varphi_{s}^{(k-1)} } \Lambda_{s}   \big)   } \bigg|_{\theta=\hat{\theta}^{(k)}} \\
	\leq & O(\f{1}{k})+ \max_t  \f{\partial d_{\varphi_t}^{(k)}}{\partial \theta}+  O\bigg(  \max_t d_{\varphi_t}^{(k)}  \bigg)
	\end{align*}
	Accordingly, we have 
	\begin{equation*}
	d_{\theta}^{(k)} \leq  O\Big(\f{1}{k} + \f{I_{k}}{\big( \sum_{t=\tau_0+1}^{k-1}  I_{t} \big)}   +  \max_t \f{\partial d_{\varphi_t}^{(k)}}{\partial \theta}+ \max_t d_{\varphi_t}^{(k)}     \Big). 
	\end{equation*}
	
\end{supplement}

\bibliographystyle{imsart-nameyear} % Style BST file

\bibliography{RobustCovid}